\documentclass[letter,scriptaddress,twocolumn, showkeys]{revtex4}
\usepackage{float} 
\usepackage{amsmath}
\usepackage{graphicx}
\usepackage{graphbox}
\usepackage{enumitem}
\usepackage{tikz}
\usepackage{tkz-euclide}
\usepackage{amssymb}
\usepackage{graphicx}
\usepackage[colorinlistoftodos]{todonotes}
\usepackage[utf8]{inputenc}
\usepackage{amsmath}
\usepackage[inline]{asymptote}
\usepackage{listings}
\usepackage{color}
\usepackage{tikz}
\usetikzlibrary{arrows}
\graphicspath{ {./images/} }
\usepackage[colorlinks=true, allcolors=blue]{hyperref}
\usepackage{setspace}

\usepackage{amscd,amssymb,latexsym,subfigure,verbatim,hyperref,epsfig,color}
 \usepackage{indentfirst}
      \newcommand{\p}{p_\text{random}}
\makeindex

\begin{document}
\title{The Power of Many: A {\it Physarum} Swarm Steiner Tree Algorithm}
\author{Sheryl Hsu $^{a}$, 
Fidel I. Schaposnik Massolo$^{b}$ 
 and Laura P. Schaposnik$^{\star, c}$ }
 \affiliation{($\star$) Corresponding author: schapos@uic.edu} 

\renewcommand{\thesection}{\Roman{section}}
\begin{abstract}
 \par We create a novel {\it Physarum Steiner algorithm} designed to solve the Euclidean Steiner tree problem. {\it Physarum} is a unicellular slime mold with the ability to form networks and fuse with other {\it Physarum} organisms. We use the simplicity and fusion of {\it Physarum} to create large swarms which independently operate to solve the Steiner problem. The {\it Physarum} Steiner tree algorithm then utilizes a swarm of {\it Physarum} organisms which gradually find terminals and fuse with each other, sharing intelligence. The algorithm is also highly capable of solving the obstacle avoidance Steiner tree problem and is a strong alternative to the current leading algorithm. The algorithm is of particular interest due to its novel approach, rectilinear properties, and ability to run on varying shapes and topological surfaces. 
\end{abstract}
\maketitle
\section{Introduction}

 \textit{Physarum Polycephalum} is a unicellular slime mold which is particularly interesting for creating algorithms due to its relative simplicity, ability to form networks, and share information through fusion \cite{Tero:2010va, plu11}. 
     In this paper, we use the model of multiple CELLs we introduced in \cite{Hsu:2021ul}, which is a cellular automaton model of {\it Physarum} organisms fusing, to form {\it Physarum} swarms. These swarms are made out of many individual {\it Physarum} organisms, allowing us to take advantage of its unique features:     
     \begin{itemize} 
     \item {\it Physarum} swarms are unique as most swarm algorithms are of more complex animals such as ants or bees while {\it Physarum} is a single-celled organisms;
    \item {\it Physarum} cells are able to solve mazes and form networks;   
     \item  {\it Physarum} cells are able to fuse and share intelligence upon merging.
   \end{itemize}
   
 We begin this paper by  recalling some background and introducing the model of multiple CELLs  \cite{Hsu:2021ul}  in Section \ref{background}. The model of multiple CELLs is built on  \cite{Gunji:2008vg}, where a {\it Physarum} organism is represented as a collection of squares on a grid where every square is either cytoplasm, cytoskeleton, or empty. Then, the key mechanism of this model is the rearrangement of cytoplasm and cytoskeleton as outside elements are introduced into the organism.

The core of the paper, where we present a novel {\it Physarum}-inspired algorithm to solve the Euclidean Steiner tree problem, appears in Section \ref{algorithm}. The Euclidean Steiner tree problem is an NP-hard problem that has been the subject of much research by both mathematicians and computer scientists since the 19th century \cite{Brazil:2014uy}. The {\it Physarum Steiner algorithm} consists of a swarm of {\it Physarum} organisms first searching for all points and fusing before finally shrinking to find the minimum Steiner tree.

  \textit{Physarum Polycephalum} typically grows in moist forests  and can be very large - up to several feet.  Biological experiments have shown that {\it Physarum} can find shortest paths, solve mazes, form high-quality networks, share information through fusion, remember past events, and adapt to its environment \cite{Nakagaki:2000wr, Tero:2010va, Boisseau:2016uz}.   In Section \ref{analysis}, we analyze the effect of cell shape and the number of cells on the algorithm before discussing the time complexity in Section \ref{time-complexity}, leading to the following findings:
\begin{itemize}
\item  \textit{\textbf{CELL shape}}. 
By changing the CELL shape:
\begin{itemize}
\item Diamond-shaped CELLs give better solutions; 
\item  Square-shaped CELLs are faster. 
\end{itemize}

\item   \textit{\textbf{CELL number}}.
Through a larger number of  {\it Physarum} organisms in the swarm:
\begin{itemize}
\item explore larger search areas,
\item find better Steiner trees, 
\item  find trees faster.
\end{itemize}
\end{itemize}

  Applications of {\it Physarum} include drug repositioning, building unconventional computer chips, approximating highways in the United States and Germany, and designing the Tokyo subway system \cite{Sun:2016tr,Whiting:2016wq, Adamatzky:2014wg,  Tero:2010va}.  In order to illustrate the novelty of the {\it Physarum Steiner algorithm} as well as the benefits of its use, we dedicate  Section \ref{applications} to describing several different uses it has:
 \begin{itemize} 
 \item \textit{\textbf{Network design}}. We use the algorithm to develop a road network in the United States and discuss characteristics  which make it particularly suited to network design and other applications. 
  \item \textit{\textbf{Obstacle-avoidance.}} We then use the algorithm to solve the obstacle-avoiding Euclidean Steiner tree problem and explain why the algorithm seems to be competitive with the current leading algorithm for this problem. 
 \item \textit{\textbf{Topological surfaces.}} We discuss the algorithm's  adaptability to varying surfaces and boundaries by considering topological surfaces such as the sphere, torus, Klein bottle, and $\mathbb{RP}^2$.
 \end{itemize}

\par We conclude this paper discussing particularly noteworthy aspects of the algorithm as well as lines of further research in Section \ref{conclusion}.

\newpage

\section{Background}\label{background}

  {\it Physarum Polycephalum} is capable of learning and remembering despite being just a single-celled organism \cite{Boisseau:2016uz}. Organisms are also able to fuse and share information with each other as they fuse \cite{plu11}. In what follows we shall first  recall the CELL model in Section \ref{CELL_model}, and then give a description of  swarm algorithms in Section \ref{swarm_algorithms} and of the Steiner tree problem in Section \ref{treeProb}.

\subsection{CELL model}\label{CELL_model}
\begin{figure}[H]
\centering \includegraphics[scale=0.35]{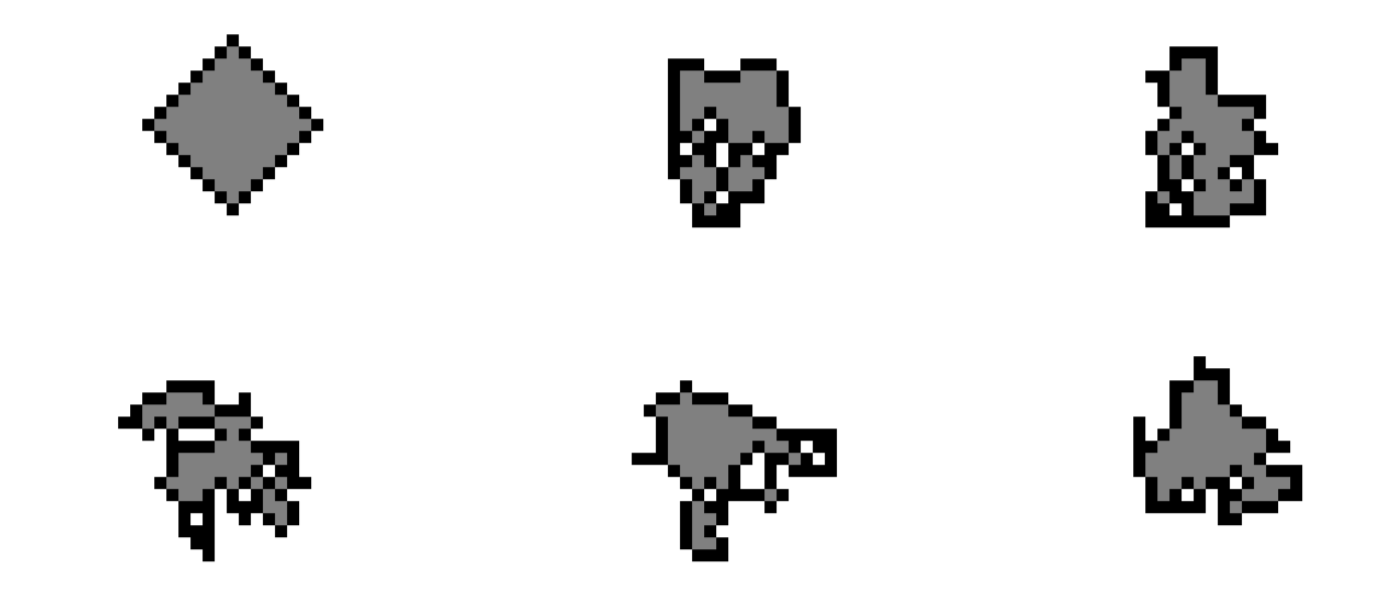} 
\caption{A CELL in amoebic motion. The black squares are cytoskeleton (state 2), the grey squares are cytoplasm (state 1), and the white squares are outside/bubbles (state 0). The CELL is initially spawned as a diamond. It then slowly reshapes itself and moves around.}
\label{fig:ameboic_ex} 
\end{figure}

The CELL model, as described in \cite{Gunji:2008vg} and expanded in \cite{Hsu:2021ul}, models a {\it Physarum} organism as a collection of squares on a grid. The key mechanism of this model is the rearrangement of cytoplasm and cytoskeleton (essentially the boundary of the cell) as outside elements are introduced into the organism. Every  square is assigned a state of $0,1$ or $2$, where $0$ represents a square that is not part of the organism,  $1$ represents a piece of cytoplasm and $2$ represents a piece of cytoskeleton. Whilst \cite{Gunji:2008vg} spawned the cells in the shape of a diamond (see Figure \ref{fig:ameboic_ex}), in the present paper we will also explore other shapes.
  \par The model is defined by an algorithm which is repeated many times: at every step, a bubble, or piece of the outside (state 0), is introduced into the organism and slowly moves throughout it. By slowly moving squares of cytoplasm many times, the CELL begins to move as a whole and take on different shapes: 
  \begin{enumerate}
  \item Randomly choose a square of state 2, which we will call the {\it stimulus point}.
  \item Choose a neighbor (a cell to the north, south, west, or east) of the stimulus point that is in state 0. Swap the states of the stimulus point and the selected neighbor. This represents the zero, which we now call a bubble, invading the cell.
  \item Replace state 1 with 2 so all squares that are part of an organism are in state 2. Set the number of moves to zero. Mark all sites as not visited.
  \item Mark the site of the bubble as visited
  \item If $s$ of the bubble's neighbors are zero, go to 8. Otherwise, go to 6.
  \item If the number of moves is larger than $n$, go to 8. Otherwise, go to 7.
  \item Randomly select one of the bubble's neighbors of state 2 and  not  previously visited. Swap the state of the bubble and the selected neighbor. Increase the number of moves by 1 and go to step 4.
  \item Reassign states 1 and 2. If a square has no neighbors that are 0, it should have a state 1. Return to step 1.
  \end{enumerate}
  
  \noindent {\bf Parameters.} In this model, there are two parameters to consider: $n$ the maximum number of swaps a bubble may take and $s$ the number of neighbors that are 0 for a bubble to stop. In the present paper, all experiments are run with $n = 1000$ and $s = 3$.
  \smallbreak
  
\begin{figure}[H]
\centering \includegraphics[scale=0.3]{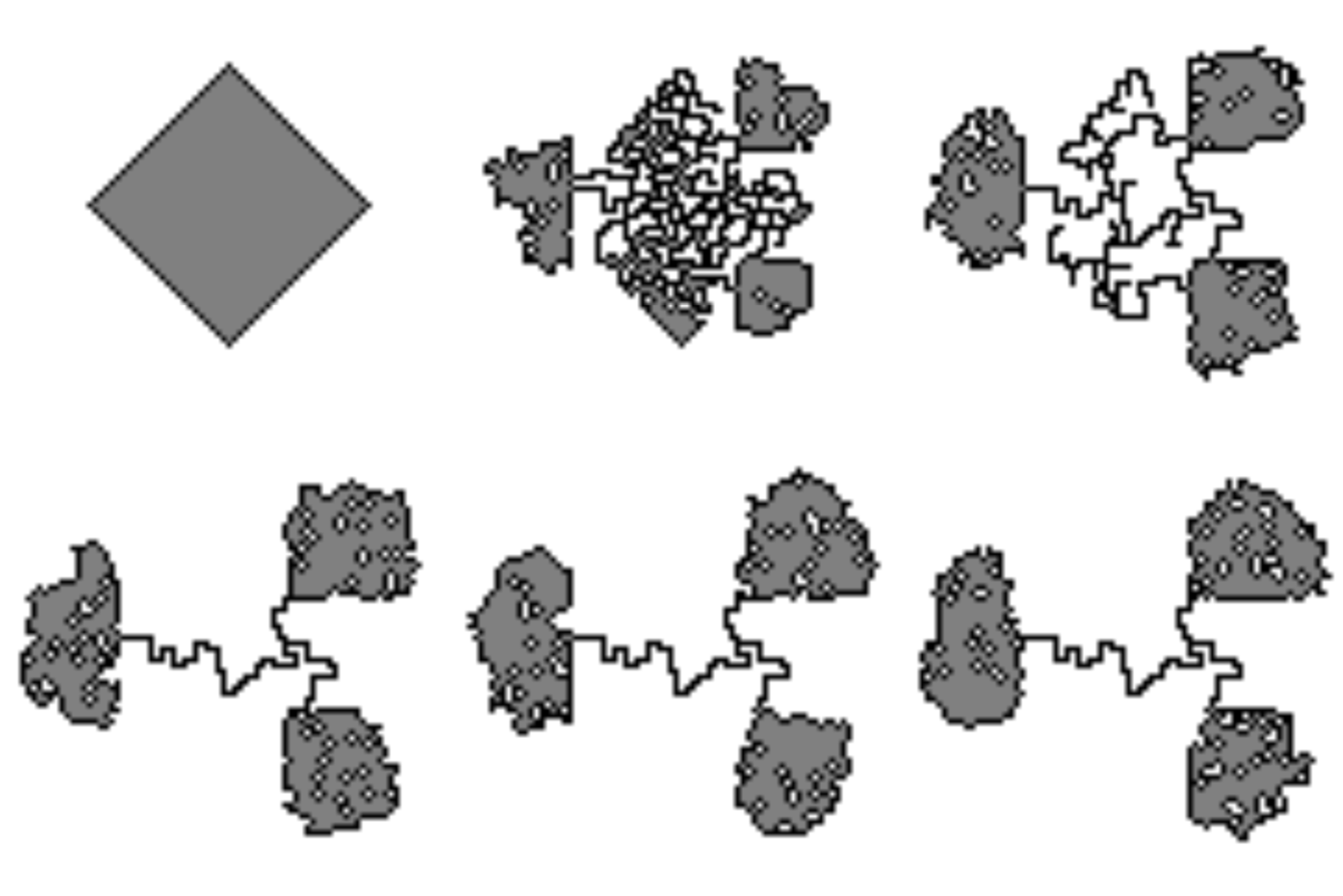} 
\caption{A CELL forming a network. The CELL is spawned as a large diamond. There are three active zones from which stimulus points are chosen. Cytoplasm is slowly moved into the active zones until a tree between the three zones is formed.}
\label{fig:tree_ex} 
\end{figure}

  \noindent {\bf Stimulus points.}  Depending on where stimulus points are selected from, the model exhibits different behaviors. As shown in \cite{Hsu:2021ul}, if stimulus points are randomly chosen, the CELL will behave like an amoeba, randomly moving around as in Figure \ref{fig:ameboic_ex}. If stimulus points are always selected from certain regions called active zones, a network between the regions forms as in Figure \ref{fig:tree_ex}. 

\begin{figure}[H]
  \centering \includegraphics[scale = 0.08]{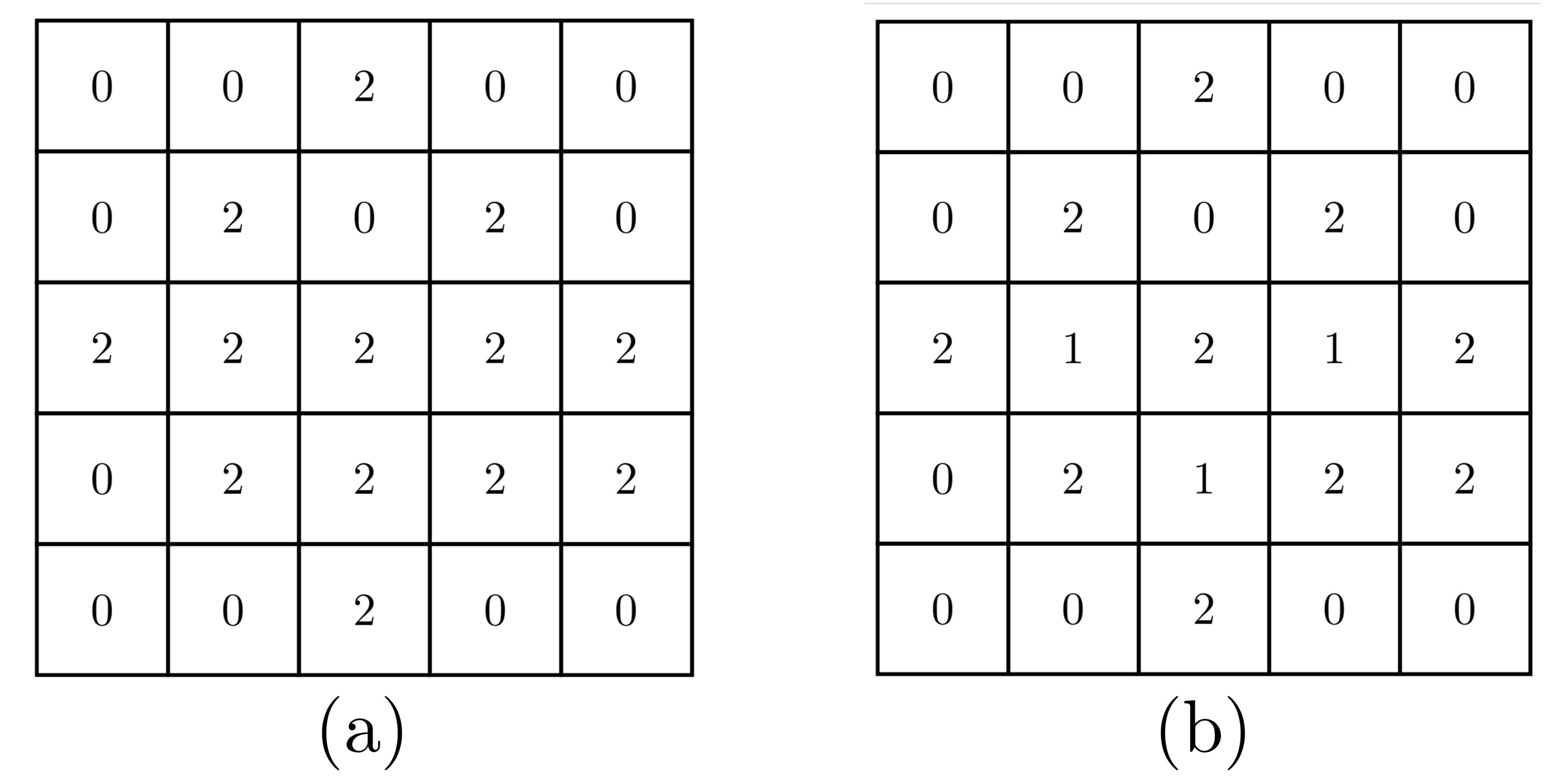}
  \caption{In (a), we continue to swap the bubble. In (b), we reassign states 1 and 2.}
      \label{box2}
\end{figure}

   \noindent {\bf Example.}   As a simple example, using the algorithm of \cite{Hsu:2021ul}  we create a small CELL  in a $5 \times 5$ grid. Following the CELL model, we start by spawning a diamond-shaped cell as shown in Figure \ref{box1} (a) below. We then randomly chose a square in state 2 to be the stimulus point (highlighted in grey). We then chose a neighbor and swap, which can be seen in Figure \ref{box1} (b):

              \begin{figure}[H]
  \centering \includegraphics[scale = 0.08]{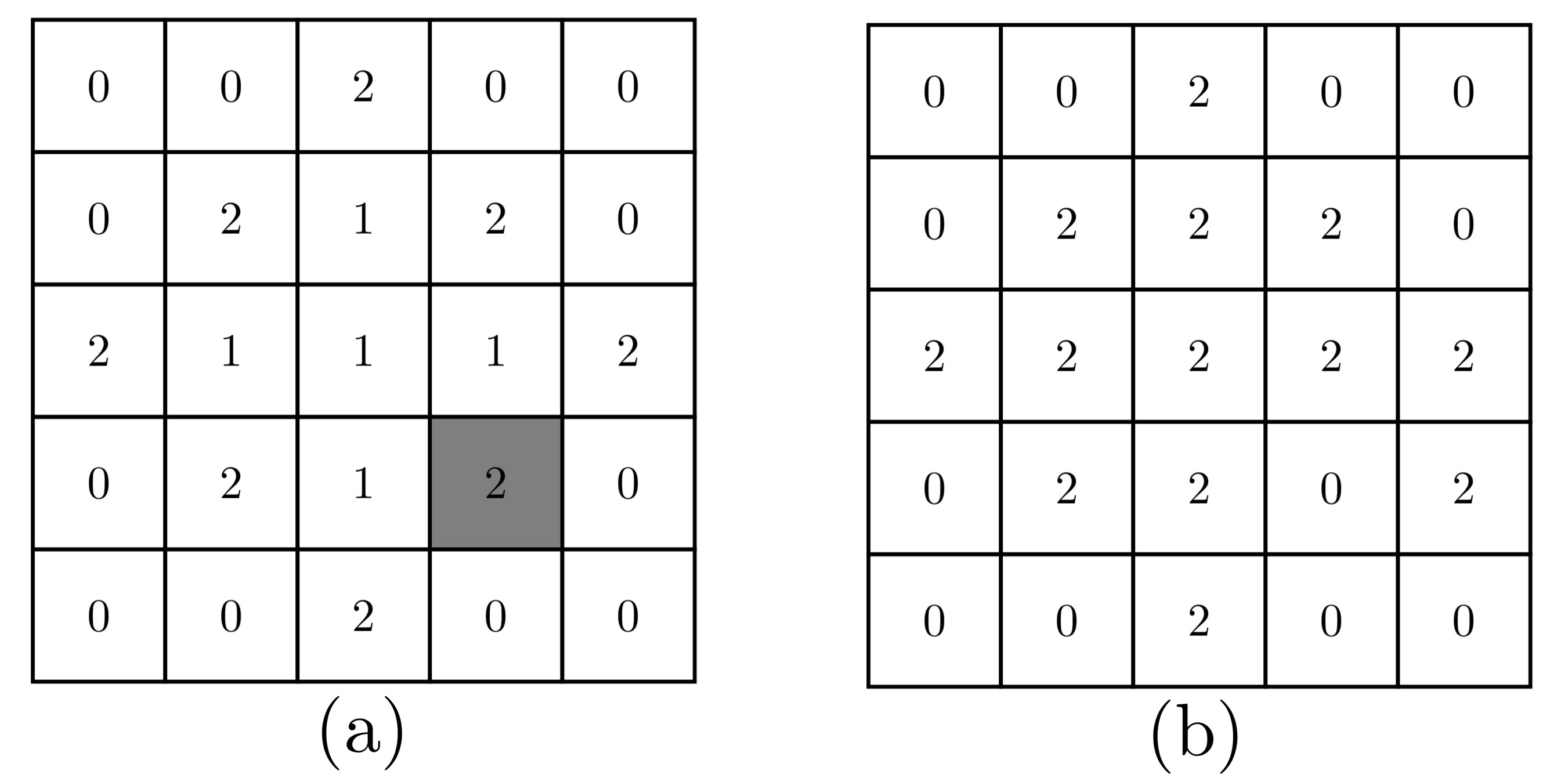}
  \caption{In (a), we select a stimulus point. In (b), we perform the first swap, introducing the bubble into the cell.}
      \label{box1}
\end{figure}
We then follow the algorithm, having the bubble randomly swap with neighbors and slowly move throughout the cell. Eventually, we reach 
 Figure \ref{box2} (a). After we stop moving the bubble, we must reassign states to the cells as shown in Figure \ref{box2} (b).
\noindent \textbf{Model of multiple cells}. The CELL model \cite{Gunji:2008vg} is extended in \cite{Hsu:2021ul} to create the model of multiple CELLs. In this model, one can spawn multiple CELLs which can be of different sizes. The main difference from the original CELL model is that stimulus points are randomly chosen from any square of state 2 across all CELLs. Once the CELLs fuse, or come in contact with each other, they are essentially treated as one CELL. Bubbles can freely move between the two fused CELLs. Each CELL is also given an ID which allows us to track which pieces of cytoplasm were originally from each cell as fusion occurs.
\par
\begin{figure}[H]
\centering\includegraphics[scale=0.22]{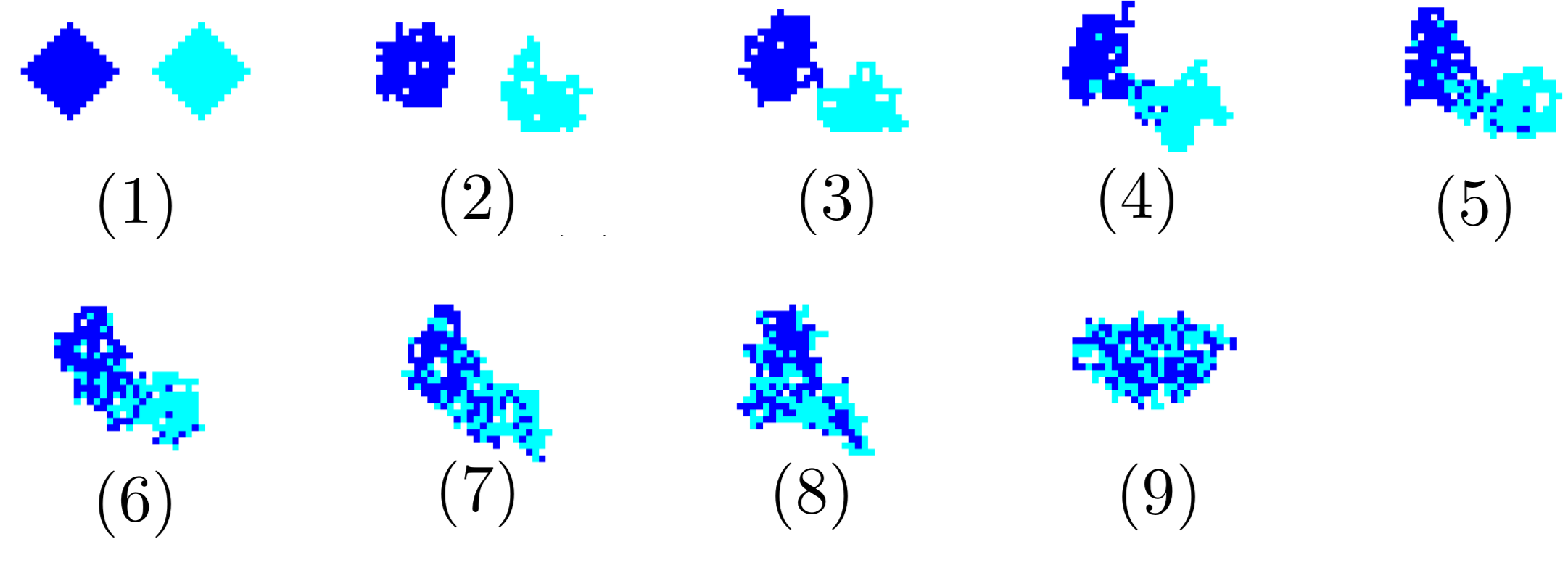}
 \caption{Two size 15 diamond-shaped cells fusing.}
 \label{fig:twoSmallCellsCombine}
 \end{figure}

 An example of  the model of multiple CELLs is shown in Figure \ref{fig:twoSmallCellsCombine}.  

\subsection{Swarm algorithms} \label{swarm_algorithms}
	Particle swarm optimization (PSO) was first introduced in 1995, and it draws heavily on biological inspiration. Like a swarm of animals, PSO consists of a swarm of agents who behave according to defined rules. For example, the velocity of an agent may be determined according to a mathematical equation that takes into account the position of other agents. Some advantages of PSO are its ability to search a large area in parallel as well as its ability to model different animal behaviors \cite{Wang:2018ve}.
	
	\par One particularly interesting PSO is the ant colony optimization (ACO). As the name suggests, this model is inspired by the behavior of ant colonies, where agents which represent ants leave behind pheromone trails which other agents can then sense and use to direct their behavior. The ACO has been used to solve the Traveling Salesman Problem and has many applications such as sequential ordering, assembly line balancing, protein-ligand docking, and DNA sequencing \cite{Dorigo:2010tf}. In what follows, we shall expand upon the model of  \cite{Hsu:2021ul} to create a novel  {\it Physarum}  swarm algorithm.   	
	
\subsection{The Steiner tree problem}\label{treeProb}
\par The Steiner tree has been a topic of great interest to mathematicians and computer scientists since the 19th century \cite{Brazil:2014uy}. Additionally, it has many practical applications such as cable routing, chip design, drug repositioning, and phylogenetic tree routing  \cite{Robins:2008wf, Caldwell:1999ts, Cho:2001tz,  Sun:2016tr, :1992ui}.
\par The general Steiner tree problem is to find the shortest tree that connects a set of given points (terminals) and can include additional points (Steiner points). There are many variations of the Steiner Tree problem. The one most relevant to this work is the Euclidean Steiner tree problem. In this problem, the goal is to find the shortest tree between a set of points on the plane. The Euclidean Steiner tree problem is a NP-hard problem. In fact, even approximating the solution within a factor of 96/95 is NP-hard, as proven by \cite{Chlebik:2008tb}. 
\par Currently, GeoSteiner is the leading publicly available Euclidean Steiner tree software. GeoSteiner has been developed since 1985 and the most recent version, GeoSteiner 5, is still relatively time consuming for large graphs - up to 24 hours for graphs with less than 10,000 terminals and up to 7 days for graphs with less than 100,000 terminals \cite{Juhl:2018th}. Since 1985, the standard approach to this problem is to first generate full Steiner trees (FSTs)  in phase 1 and then combine a subset of the FSTs to find a minimum Steiner tree in phase 2 \cite{Winter:1985vb}. 
\par Finally, there is  a variation of the Euclidean Steiner tree problem which shall be of interest in the present paper:  the obstacle avoiding Euclidean Steiner tree problem where the tree needs to avoid certain regions of the plane. The leading obstacle avoiding Euclidean Steiner tree algorithm \cite{Zachariasen:1999tv} was developed in 1999 and has a run time of multiple hours for graphs with only 150 terminals.

\section{Physarum Steiner Algorithm}\label{algorithm}

In what follows we shall create the {\it Physarum Steiner algorithm},  which uses the model of multiple CELLs to solve the Euclidean Steiner tree problem. An implementation of this algorithm has been made available at \cite{Hsu:wo}. Whilst this algorithm is a swarm algorithm inspired by those described in Section \ref{swarm_algorithms}, there are some important differences we should note:
\begin{itemize}
\item  While many swarm algorithms model complex animals like ants, the {\it Physarum} swarm uses a simple unicellular organism. This results in larger, simpler swarms which may scale better. 
\item {\it Physarum} posses many unique abilities such as maze solving and network generating.
\item In addition, the CELLs in the {\it Physarum} swarm have the ability to share intelligence through fusing.
\end{itemize}
\par
\begin{figure}[H]
\centering \includegraphics[scale=0.2]{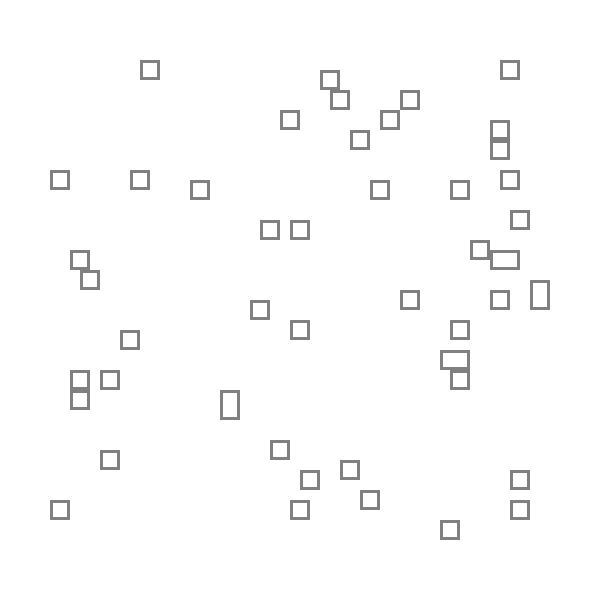}
\caption{Sample grid, with active zones represented as 2x2 squares. Active zones next to each other are simply represented as a rectangle visually although it does not make a difference in the algorithm.}
\label{fig:sample_active_zones}
\end{figure}

 This algorithm has two segments: first {\it foraging} where the CELLs find all points and then {\it shrinking} where the CELL looses cytoplasm as it finds the minimum Steiner tree. We represent the terminals, or points we want to find a Steiner tree for, as 2$\times$2 squares (active zones) on a grid as shown in Figure \ref{fig:sample_active_zones}. In this model,  $N$ is the number of points and $M$ is the length of the square edge of the grid (most of the grids are square although we will discuss grids of other shape later).

\noindent {\bf Foraging section.} We take all of the mechanics from the CELL model - the movement of bubbles and assignment of states stays the same. We mainly add a more complex selection of the stimulus point to cause the organism to form a Steiner tree. 
  Initially, we spawn a {\it Physarum} swarm. This consists of as many CELLs as we want. Let $L$ be the number of CELLs initially spawned. We then begin to run the CELL algorithm with some modifications. At every iteration, we keep track of the number of active zones/points the CELLs are currently in contact with. Call this number $a$. We also keep track of the number of disjoint CELLs which contain at least one point. We call this number $b$. 
\par The stimulus point can be chosen from any square of cytoplasm in state 2; it is not limited to a certain CELL. In every iteration, we have two options for the stimulus point. We can choose a piece of cytoplasm inside an active zone that has already been found, which will bring cytoplasm to the active zone and help prevent the CELLs from moving away from active zones that have already been discovered. The other option is randomly choosing any square with state 2, as in the original CELL model. This helps the CELLs explore in random directions and find more active zones. The probability $\p$ that we choose the second option (random) is defined according to:
\begin{equation}
\p = \frac{N - a + b - 1}{N + L}.
\label{eq:non_active}
\end{equation}
\par In \eqref{eq:non_active},    the number of points not found is represented by  $N - a $, and  $b - 1$ represents the number of cells that still need to fuse. As a result, the more points that are left to find and the more cells that need to still fuse, the higher the probability of choosing a random stimulus point, which favors exploration.

Then, with probability $\p$, we choose a random stimulus point. As in the original CELL model, we simply choose any square that has a state of 2. 
With probability $1 - \p$, we choose a stimulus point from inside an active zone. To do this, we loop through all active zones and randomly select a square of state 2 that is inside a zone. If there are no squares of state 2 inside an active zone, we will instead choose a random stimulus point. 
  When $\p$ becomes zero (one CELL organism is connecting all points), it is time to move from the {\it foraging} portion of the algorithm to the {\it shrinking} one.
\begin{figure}[H]
\centering \includegraphics[scale=0.2]{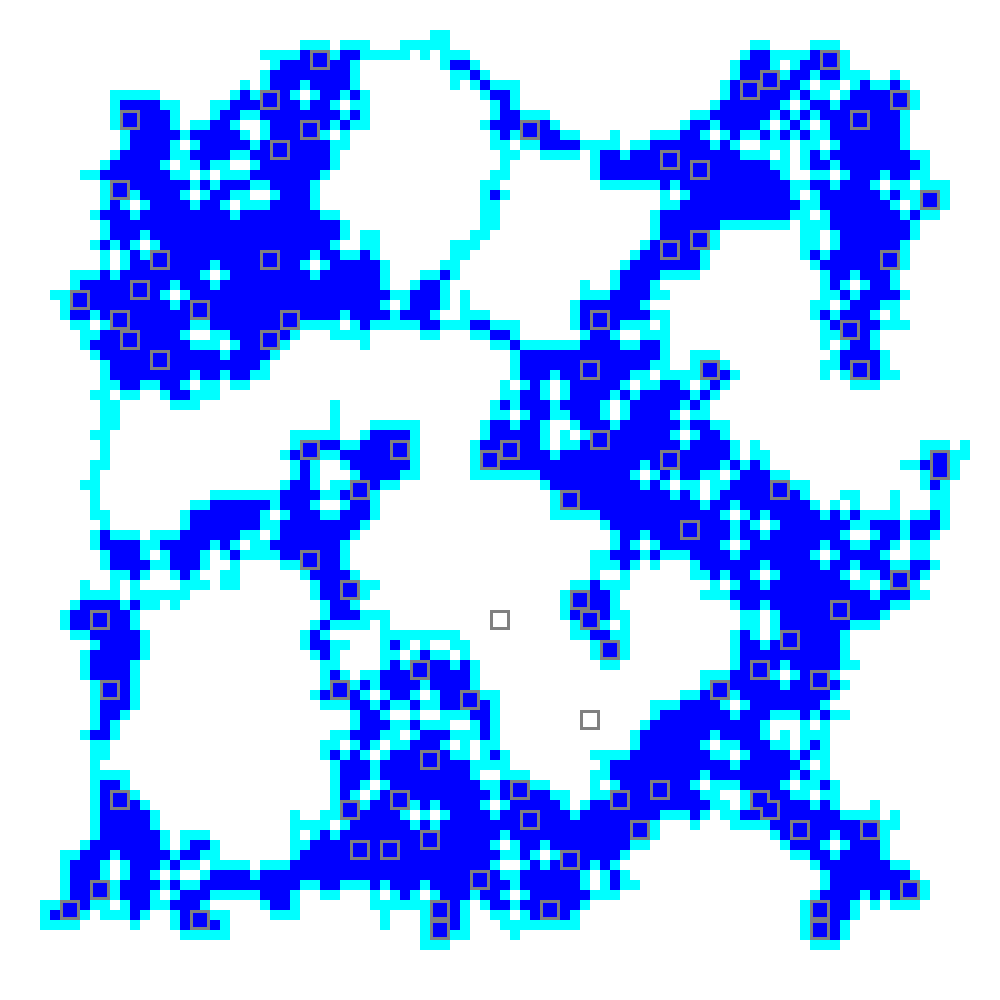}
\caption{The algorithm is currently in the foraging state. The CELL needs to discover the last two active zones before entering the shrinking phase.}
\label{fig:sample_foraging}
\end{figure}

\noindent {\bf Shrinking section.}   In this portion of the algorithm, $\p$ is zero so we are always selecting stimulus points in active zones. However, when there are no possible stimulus points in active zones, we randomly choose a piece of cytoplasm to remove from inside an active zone. We change the state of that square to zero, decreasing the mass of the CELL and also creating some viable stimulus points. We also keep track of the number of iterations since the area of the cell last changed. When this number passes a threshold (1 million was used in this paper), the algorithm terminates. 
\begin{figure}[H]
\centering \includegraphics[scale=0.15]{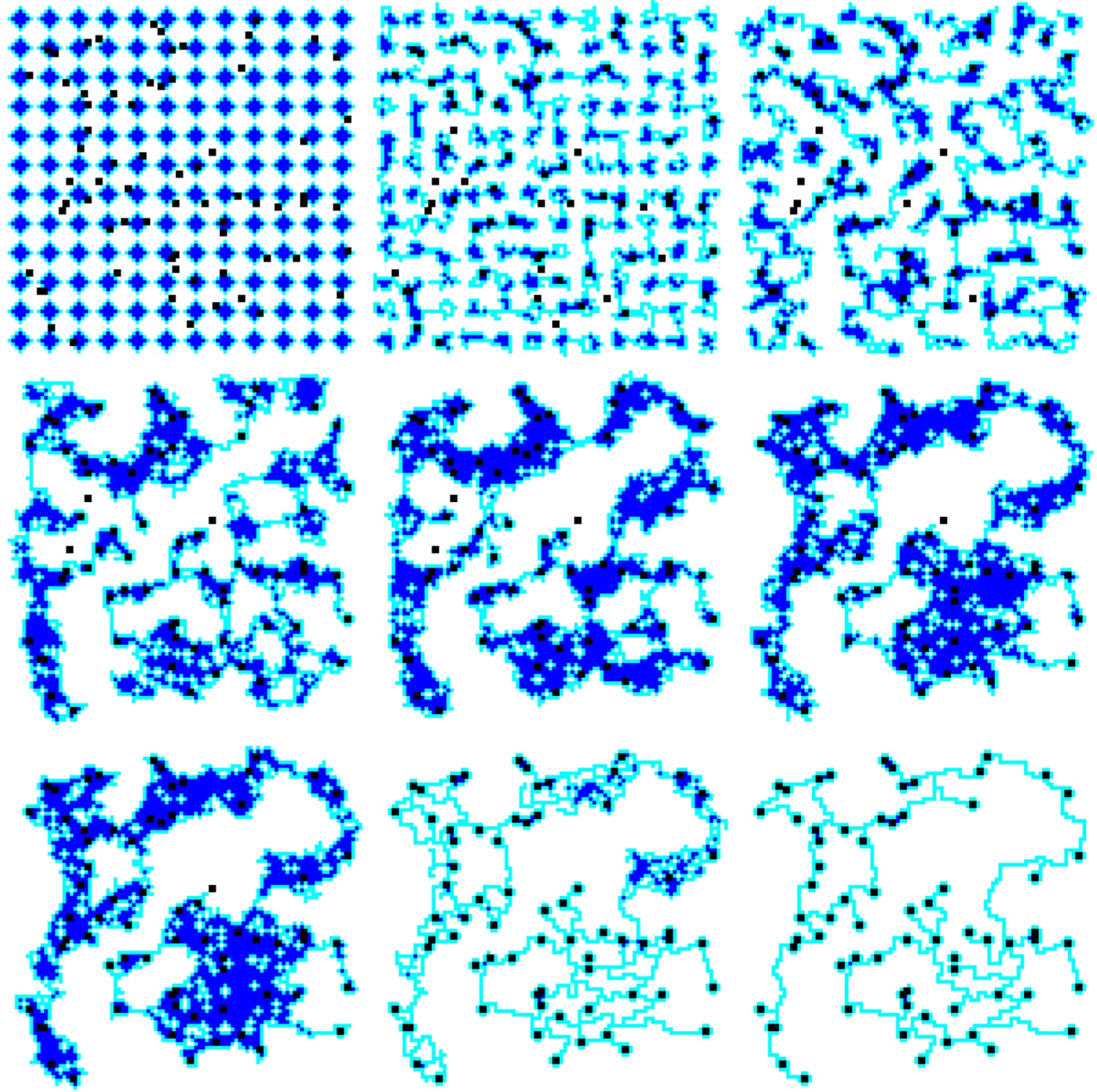}
\caption{Progression of the algorithm:  We begin by spawning 144 CELLs of size 7 on the grid in the top left image. Throughout the top two rows, we can see the {\it foraging} phase occur. The cell slowly finds all the points and fuses together into one big cell. In the bottom row, we can see the {\it shrinking} phase. The CELL slowly loses cytoplasm, with the final result being displayed in the last image.}
\label{fig:sample_full_algo}
\end{figure}

\section{CELL shape and number}\label{analysis}
In what follows we shall consider how the different shapes of CELLs, as well as their number, influence the results one may obtain through the {\it Physarum Steiner algorithm} described in Section \ref{algorithm}.

\subsection{CELL shape}
\par
\begin{figure}[H]
\centering \includegraphics[scale=0.11]{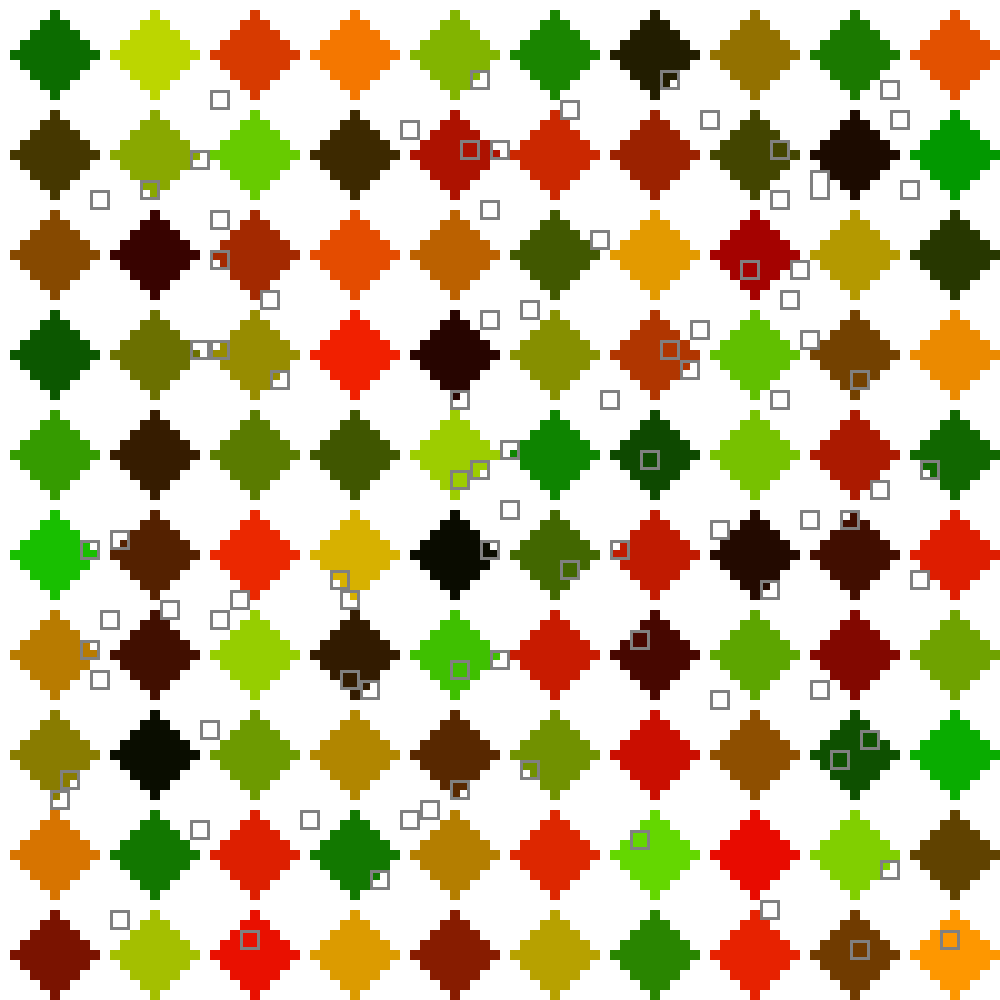} \hspace{1pt} \includegraphics[scale=0.1]{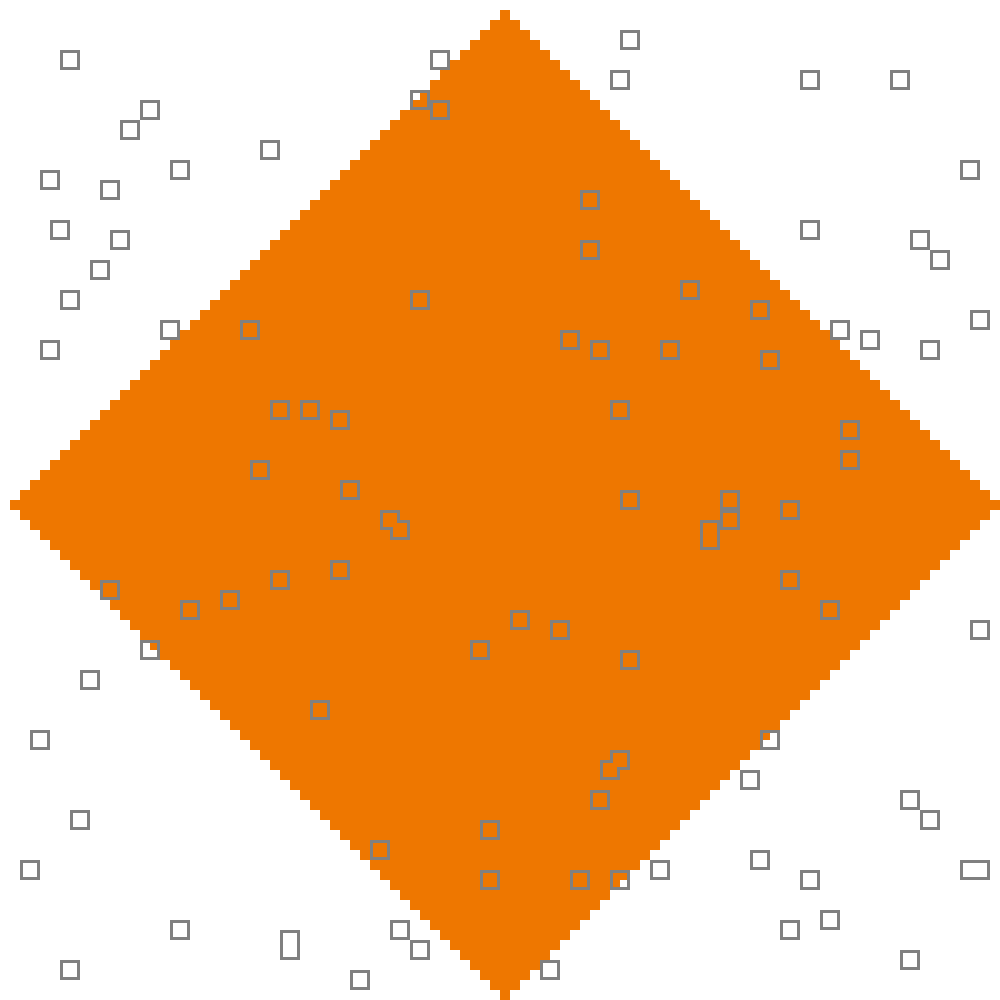} \\ \vspace{4pt}
\centering \includegraphics[scale=0.11]{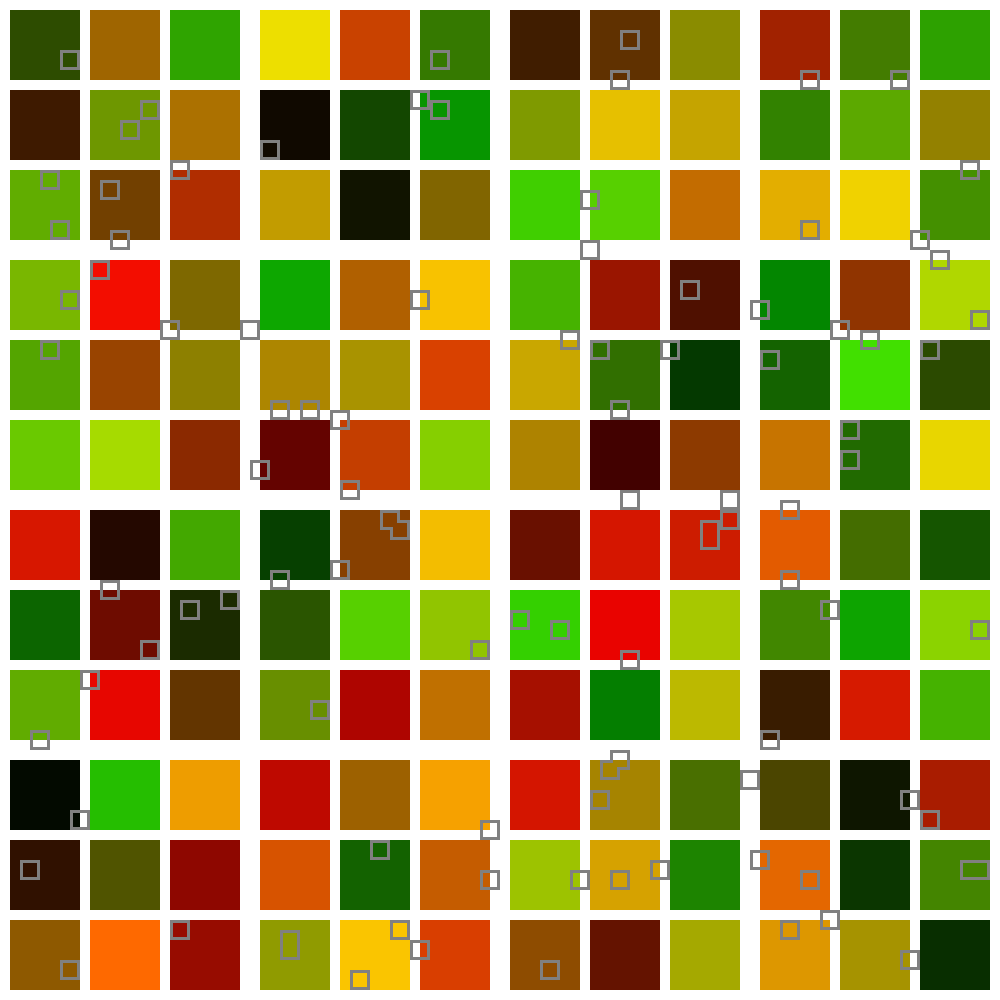} \hspace{1pt} \includegraphics[scale=0.1]{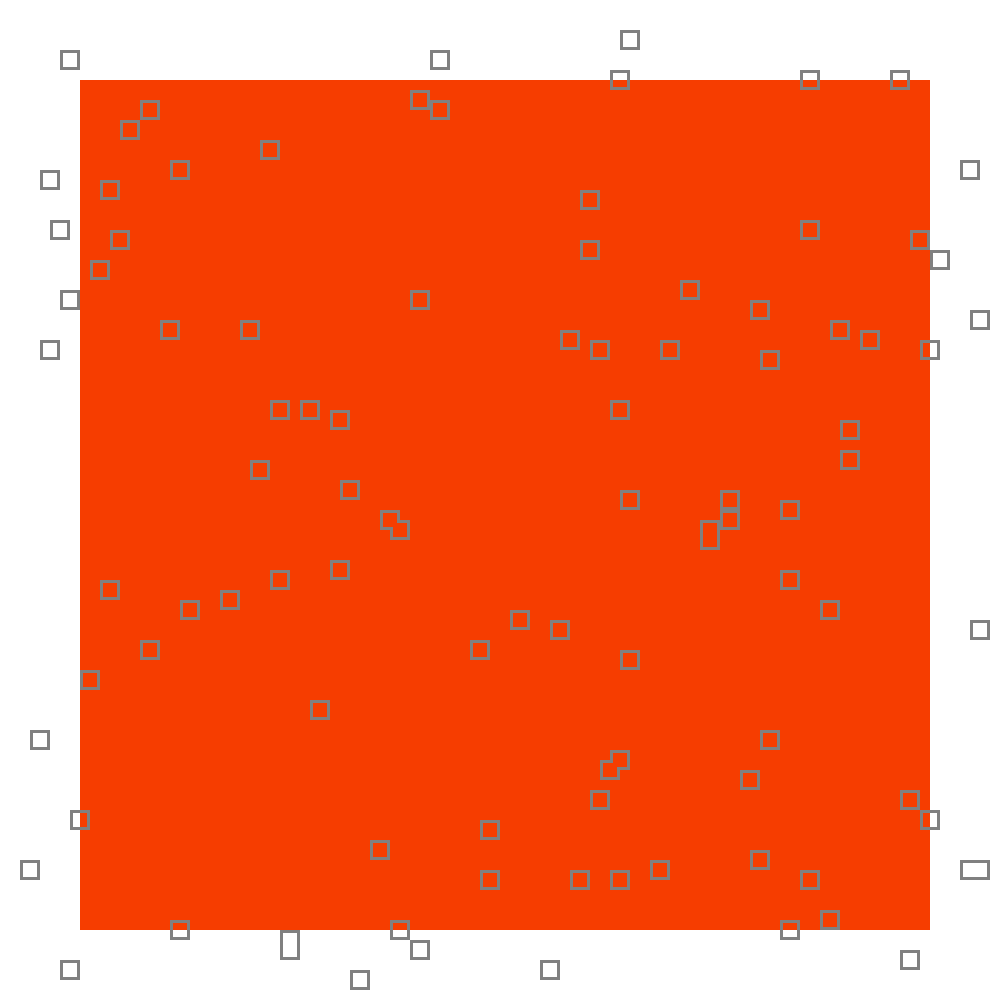}
\caption{In the top left are many diamond shaped cells. On the top right is one big diamond shaped cell. On the bottom left are many square shaped cells. On the bottom right is one big square shaped cell.}
\label{fig:square_diamond}
\end{figure}

 Although  \cite{Gunji:2008vg}  and \cite{Hsu:2021ul} considered diamond shaped CELLs, we shall consider here CELLs with other shapes. In particular, we shall study square CELLs, since they provide multiple advantages: \begin{itemize}
\item There is more cytoplasm in general since squares are more compact than diamonds (see Figure \ref{fig:square_diamond});
\item  In addition, it is easier for adjacent cells to fuse;

\item Since the search space is a square, square cells are able to cover the space much more optimally, especially for large cells (consider one large diamond versus one large square); 
\item Finally, square CELLs can be any size while diamond CELLs could only be odd. This allows us to gather more data points with different sizes.
\end{itemize}

\begin{figure*}[t]
\centerline{
\hspace{-7in} 
\includegraphics[scale=0.075]{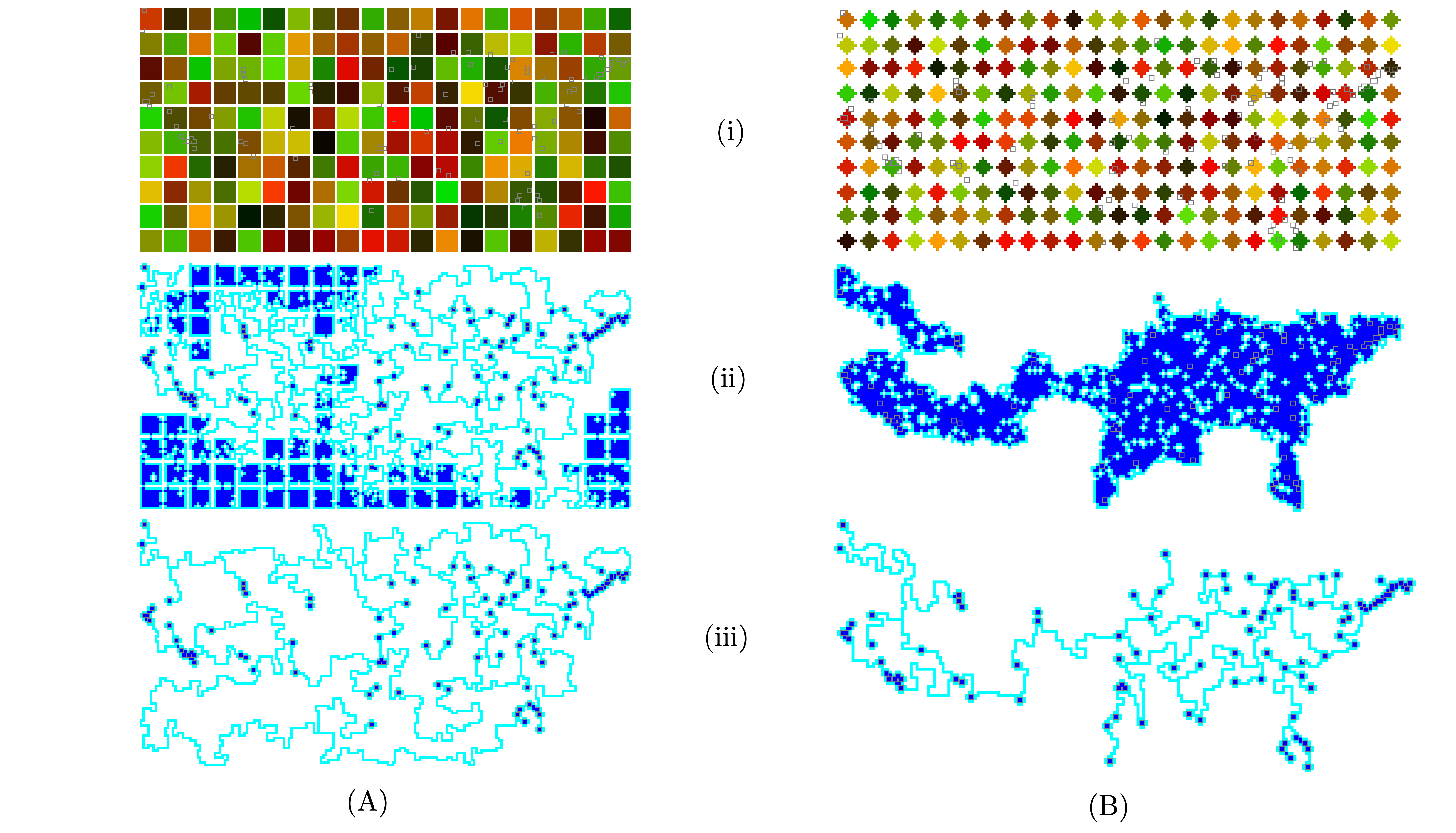} }
\caption{{\bf (A).} Size 9 squares spawned one apart. Bottom image is final solution. {\bf (B).} Size 7 diamonds spawned one apart. Bottom image is final solution.}
\label{fig:bad_squares}
\label{fig:good_diamonds}
\end{figure*}

Through  simulations, we see that diamond cells often times produce better solutions. When squares are packed tightly together, there is much more cytoplasm, which results in very little time in the {\it foraging} shape and a lot in the {\it shrinking} phase, which can result in worse solutions. 
\smallbreak
\noindent {\bf Example A.} In order to illustrate the above,  in Figure \ref{fig:bad_squares}, we begin with squares tightly packed, which have a lot of cytoplasm. Since the squares are so tightly packed (1 apart), all points are found very quickly, for if any piece of cytoplasm in a square is moved, it will lead to a connection with a neighboring cell. As a result, a lot of squares are connected and part of the network, even if they are not close to any of the points as shown in the middle image of Figure \ref{fig:bad_squares}. Shrinking these extra squares takes a long time and can also result in long paths which are far out of the way as shown in the bottom image of Figure \ref{fig:bad_squares}.

\bigskip

 \bigskip

{\bf Example B.} In contrast to Example A, in Figure \ref{fig:good_diamonds} we consider diamond-shaped cells. We see that the CELLs are initially diamonds with less overall cytoplasm. The CELL then spends quite a few iterations in the {\it foraging} phase. Although this does take time, it allows the cytoplasm to move towards a centralized location around the active zones. When the CELL finally proceeds into the {\it shrinking} phase, there is less cytoplasm to remove and no out of the way paths. The downside to this is the increased time which in some cases can be very long (over 100 million iterations) or in some cases never finish.
\par Although it is possible to simply spawn smaller squares and increase the spacing between them, it becomes difficult for the squares to merge. In contrast, the shape of the diamonds allows for there to be less cytoplasm why still having cells that merge relatively quickly.

\subsection{The effect of multiple CELLs}
In this section, we examine the effects of the number of cells we use. We run 10 trials on 10 grids for a total of 100 trials on every cell size and number of cells. For each trial, we measure the total amount of cytoplasm that is initially spawned which we will henceforth refer to as area. From there, we are able to compute the search area as a percentage of cytoplasm. By search area, we mean the number of squares in the grid (for example a $100 \times 100$ grid will have search area 10000). 

\noindent {\bf Success rate.} One of the most important characteristics is the proportion of runs that are successful. Sometimes, this algorithm is unsuccessful. The CELLs may miss a point early on and move far away from that point, making it almost impossible to ever find that point in some cases. There may simply not be enough cytoplasm for two far away CELLs to fuse into one.  For each number of cells (1, 9, 25, 100), we try various sizes/amounts of cytoplasm and compute the proportion of trials (out of 100) that successfully terminate within 10 million iterations. 
\begin{figure}[H]
\centering \includegraphics[scale=0.5]{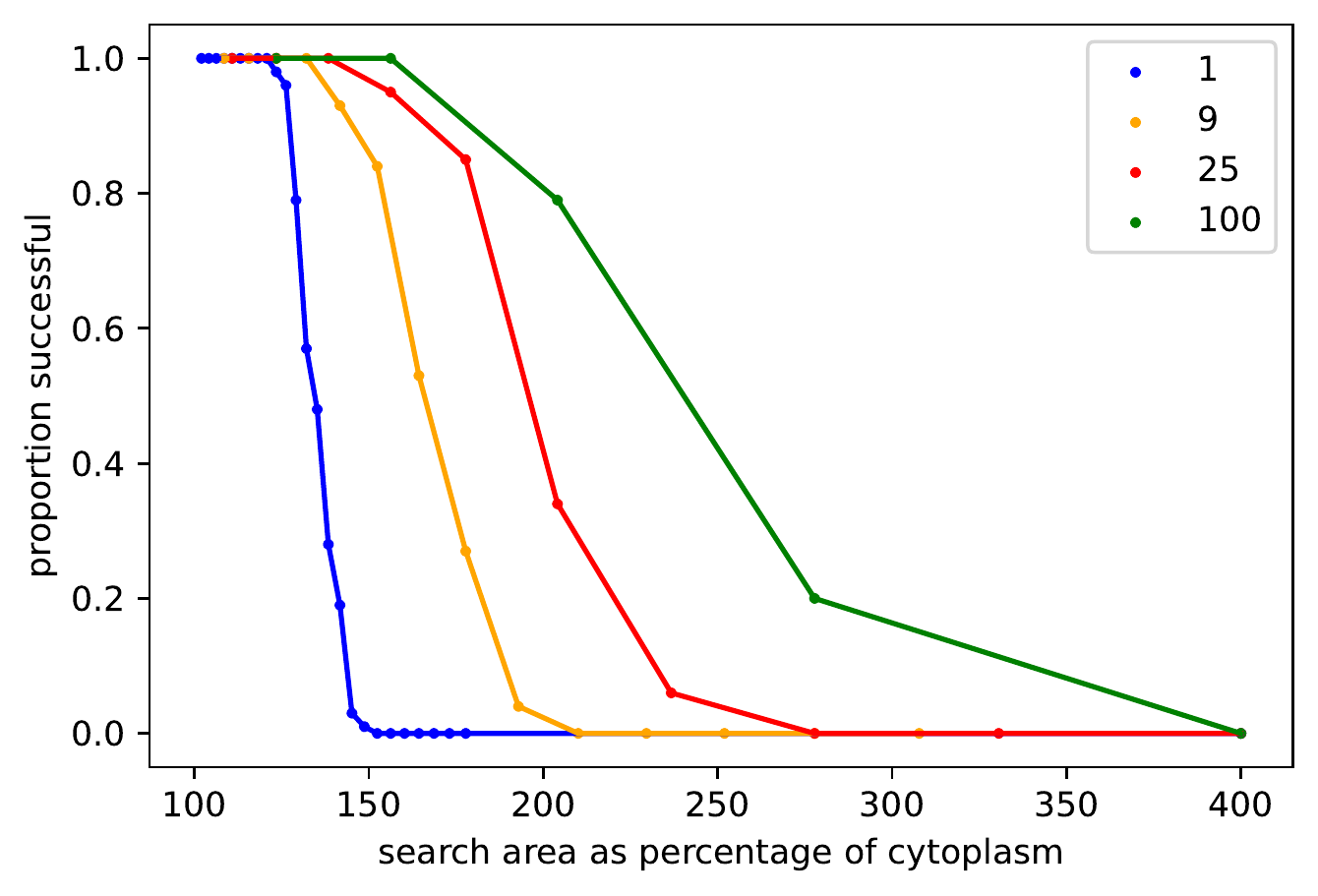}
\caption{This graph shows the proportion of trials that are successful versus the search area as a percentage of cytoplasm for trials with 1, 9, 25, and 100 cells.}
\label{fig:200_proportions}
\end{figure}

In Figure \ref{fig:200_proportions}, we see that the green line (100 cells) extends much further to the right than the blue line (one cell). Thus, the more cells there are, the larger of a search area we can explore. This is mainly because with more cells, we can spread out our cytoplasm instead of having to have it concentrated in certain areas.
\smallbreak

\noindent {\bf  Solution length.} Another important metric to consider is the solution length. We measure how good the solution is by counting the amount of cytoplasm when the algorithm terminates. We ignore any cytoplasm that is part of a disjoint CELL that does not contain an active zone, or in other words is separate from the CELL that actually forms the tree. 
\begin{figure}[H]
\centering \includegraphics[scale=0.55]{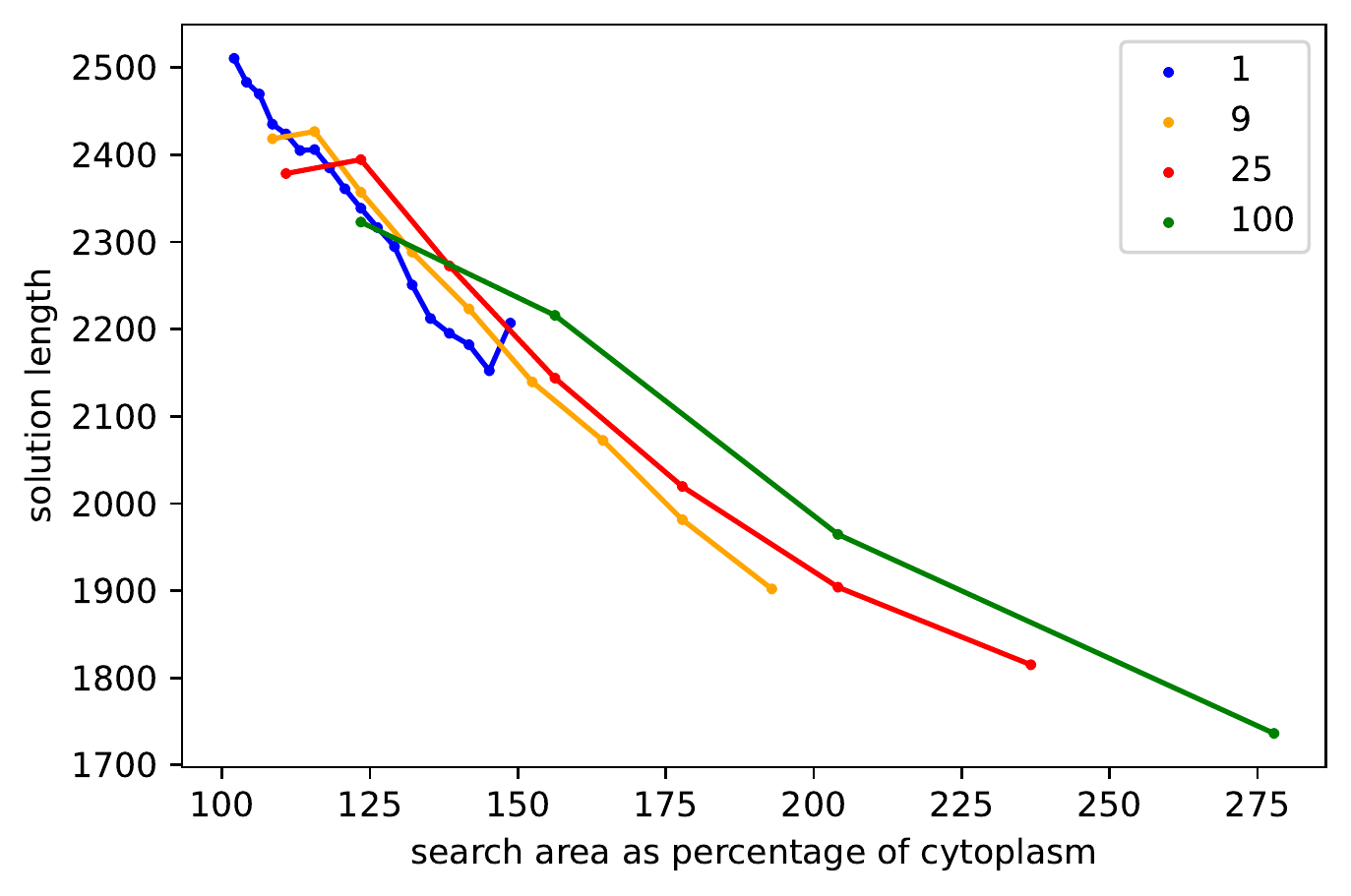}
\caption{This graph shows the length of solutions versus the search area as a percentage of cytoplasm for trials with 1, 9, 25, and 100 cells. Note that we exclude unsuccessful runs to produce this graph.}
\label{fig:200_length}
\end{figure}
In Figure \ref{fig:200_length}, we see that as the search area as percentage of cytoplasm increases, the solution gets better. This is probably because there is comparatively less cytoplasm to begin with. In addition, we see that as the number of cells increases, it is possible to find a better solution. This is probably because using more cells allows us to still find solutions even with less cytoplasm, which results in better solutions. Trials with 100 cells found the shortest solutions (rightmost data point).

\smallbreak
\noindent {\bf Run time.} The last metric we consider is the run time. We consider the true number of iterations the algorithm runs for. 
\begin{figure}[H]
\centering \includegraphics[scale=0.55]{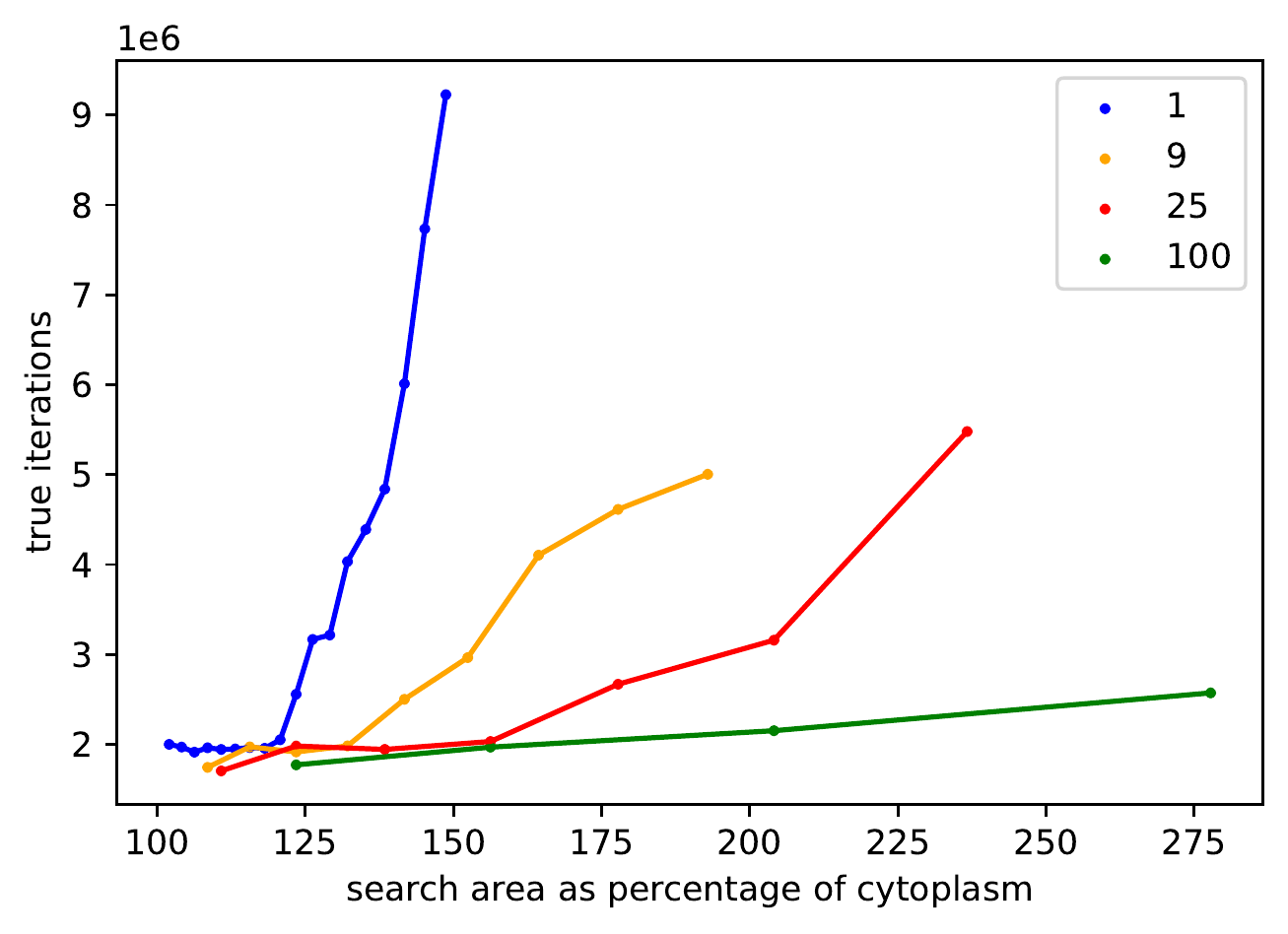}
\caption{This graph shows the number of iterations as a percentage of cytoplasm for trials with 1, 9, 25, and 100 cells. Note that we exclude unsuccessful runs to produce the graph.}
\label{fig:200_t_iterations}
\end{figure}

By true iterations, we account for the fact that in a parallel algorithm or real-world {\it Physarum} organism, multiple CELLs would be introducing and moving bubbles at the same time. As a result, every iteration we add $\frac{1}{\text{number of disjoint CELLs}}$ to the true iteration count. In Figure \ref{fig:200_t_iterations}, we see that the more CELLs there are, the lower the number of iterations. This may be because with more cells, the cytoplasm is more spread out and therefore there are less out of the way points which may take a very long time to find. 
\par From the above analysis, we see  that using more CELLs allows us to explore bigger search areas, find shorter solutions, and solve problems faster.

\section{Time complexity}\label{time-complexity}
In what follows we will analyze the time complexity of the {\it Physarum Steiner algorithm}. There are two variables we must consider: $N$ the number of points and $M$ the size of the grid. We first independently analyze $N$ and $M$ then vary $M$ at a fixed ratio to $N$. We measure the number of iterations that the algorithm takes to terminate. It is important to note that each iteration of the {\it Physarum Steiner algorithm} is not necessarily linear, but this is very dependent on the specifics of the implementation, which is beyond the scope of this paper. We used size 9 square cells spaced one apart which led to a short foraging phase and a much longer shrinking phase.
\par
\begin{figure}[H]
\centering \includegraphics[scale=0.5]{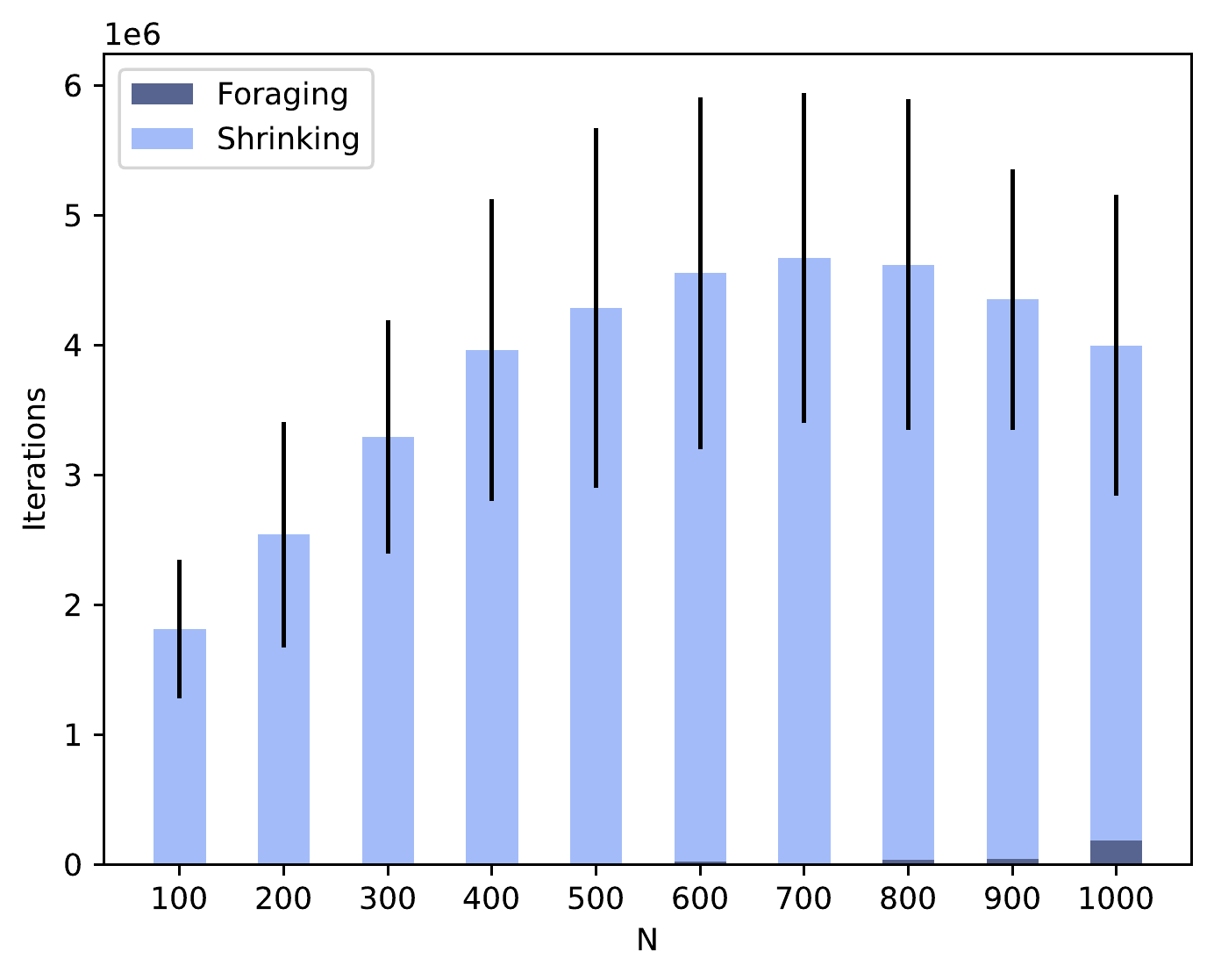} 
\caption{Iterations varying with $N$. Black lines on bars represents error (standard deviation). One failed $N=1000$ trial is excluded from graph shown.}
\label{fig:N_time}
\end{figure}

 
  We first analyze the time complexity in terms of $N$, the number of points. We set $M$ to be a constant of $100$. For every value of $N$ from $100$ to $1000$, we generate 10 random $100 \times 100$ graphs. We run $10$ trials on each of the graphs, for a total of $100$ trials for every value of $N$. We have a very high success rate. In particular, $N = 1000$ had one trial fail to complete within 10 million iterations, leading to a success rate of $0.99$. This failed trial is excluded from the iteration and time graphs in Figure \ref{fig:N_time}. All other values of $N$ had a 100 percent success rate. 

\par In Figure \ref{fig:N_time}, we see that the number of iterations appears to initially increase before decreasing. We hypothesize that the number of iterations decreases for larger values of $N$ because as $N$ increases, the final solution gets longer and thus there is not as much cytoplasm that needs to be removed through shrinkage. In addition, since there are more points, pieces of cytoplasm are more likely to be close to a point. Since cytoplasm is removed at points, or in other words bubbles are propagated from the points, if cytoplasm is closer to points there is a higher probability that it will be removed. The time complexity of this algorithm with regards to $N$ is thus less than linear. This is noteworthy considering the runtime and time complexity of other Steiner tree algorithms.
\\\\
\par We now analyze the time complexity in terms of $M$, the size of the grid. We set $N$ to 100 and $M$ takes on values from $50$ to $250$. As before, we run $10$ trials on $10$ graphs for every value of $M$. All trials in this experiment were successful.

\begin{figure}[H]
\centering \includegraphics[scale=0.5]{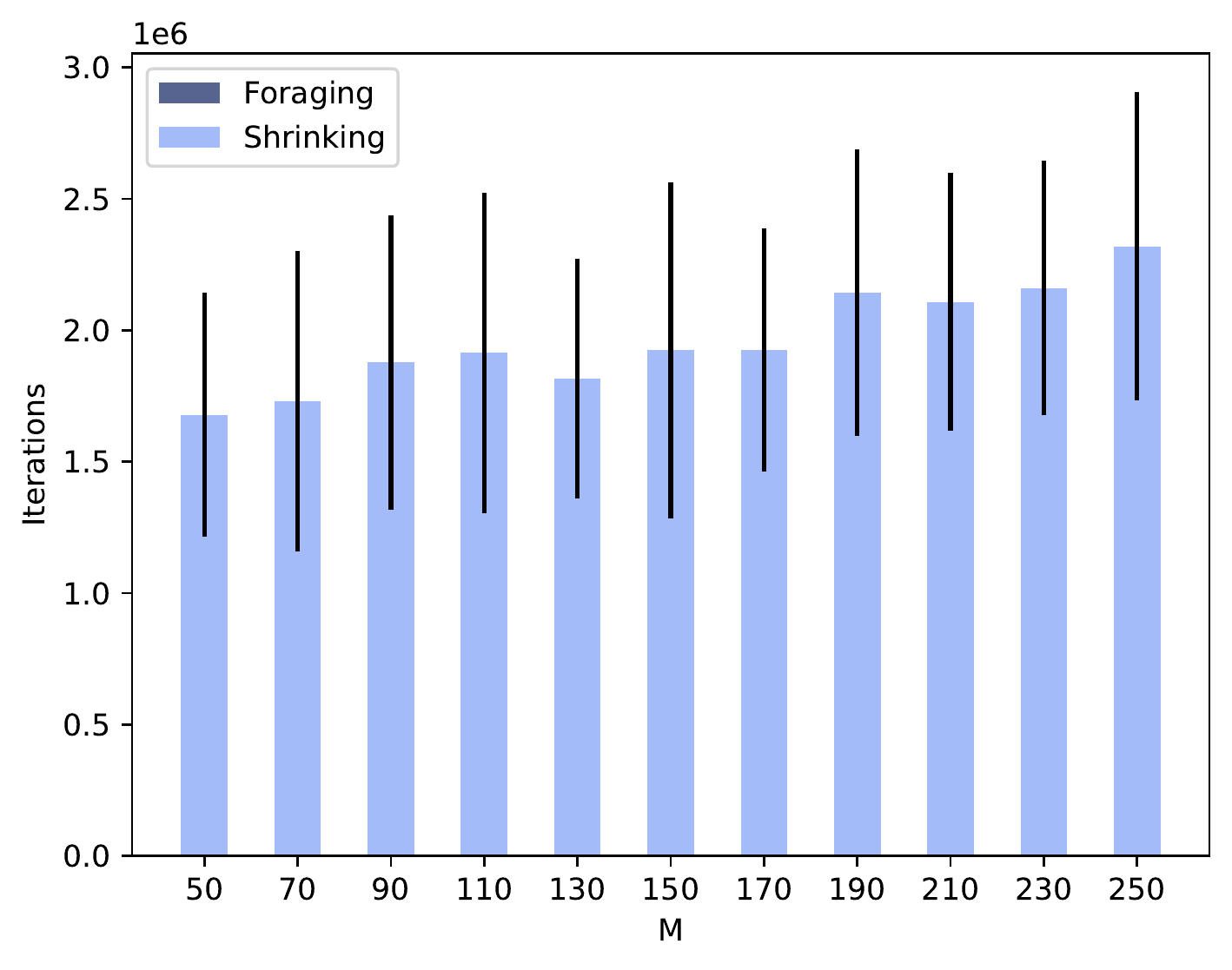} 
\caption{Iterations varying with $M$. Black lines on bars represents error (standard deviation).}
\label{fig:M_time}
\end{figure}
\par In Figure \ref{fig:M_time}, we see that the number of iterations appears to very slowly increase roughly linearly. The low slope of this line (less than $10^6$ iterations for a 200 unit increase in $M$) suggests that this algorithm scales well to larger search areas.

Finally, we shall consider what happens when $N$ varies with $M$. We set $N$ to always be two percent of the search area, or $M \times M$. We run trials where $N$ takes values from $100$ to $1000$ and $M$ is computed according to Eq.~\eqref{eq:M_2_percent}. 
\begin{equation}
M =\left[ \sqrt{\frac{N}{0.02}} \right]
\label{eq:M_2_percent}
\end{equation}
\par As before, we run 10 trials on 10 graphs for every value of $N$.  $N=500$ and $N=700$ had a success rate of 97 percent and $N = 1000$ had a success rate of 99 percent. All other trials were successful.
\begin{figure}[H]
\centering \includegraphics[scale=0.5]{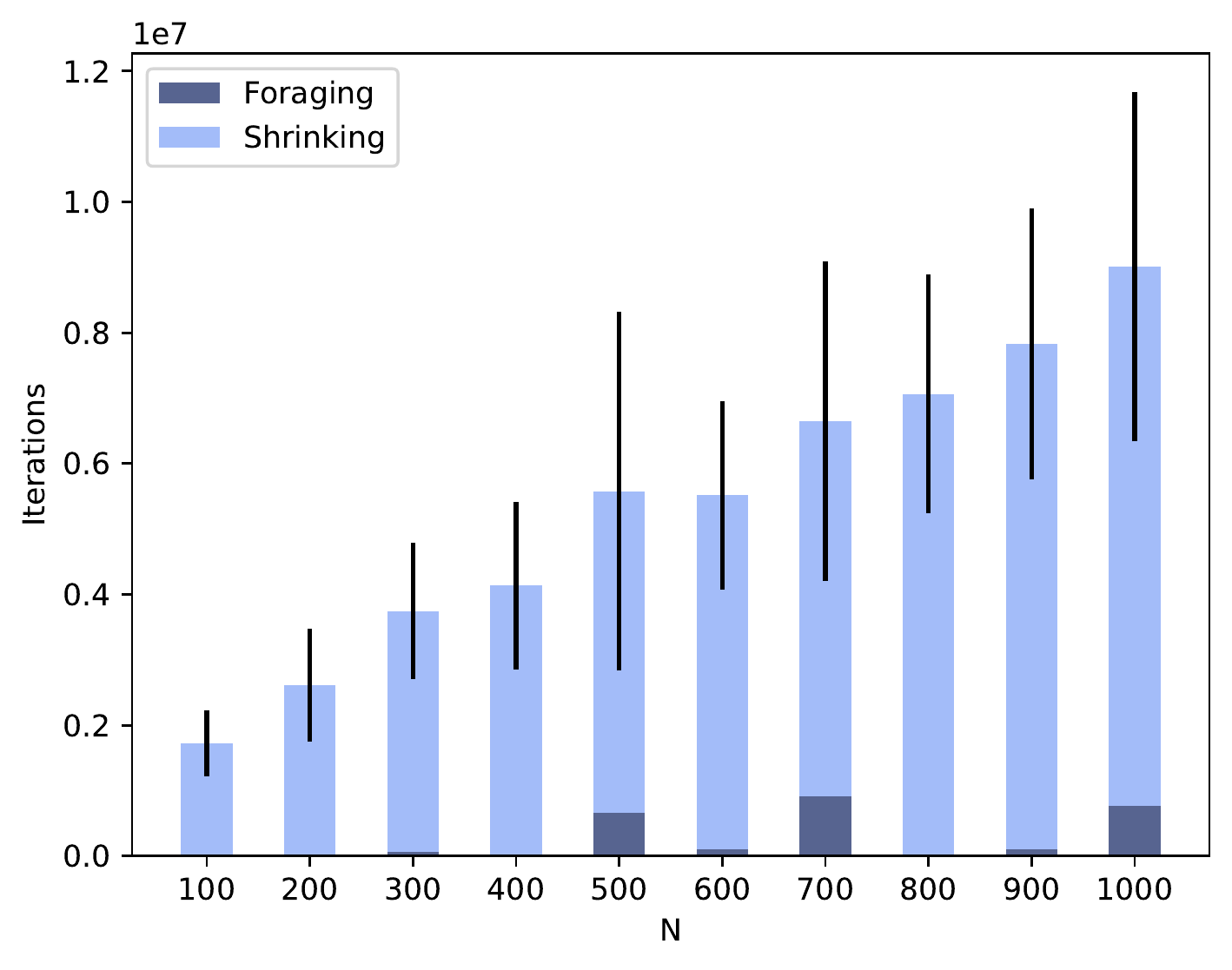} 
\caption{Iterations varying with $N$. Black lines on bars represents error (standard deviation). Failed trials are excluded from both graphs.}
\label{fig:scale_time}
\end{figure}
\par In Figure \ref{fig:scale_time},  we see that the number of iterations increases linearly with $N$. We see some values of $N$ that have a considerable amount of foraging. This may be due to the random generation of grids resulting in grids with points that are concentrated in out of the way locations.
\\\\
\par In summary, we can think of $M$ as a control of how fine or detailed the solution will be. Adding more points ($N$) increases the run time by a less than linear factor and increasing the search area $M$ comes at a linear factor. With a linear time complexity, we believe that the algorithm will scale well to large problems.
\section{Applications} \label{applications}
As mentioned before, the behaviour of  {\it Physarum} and the models it has inspired have found many different uses,  among which are drug repositioning, building unconventional computer chips, approximating highways in the United States and Germany, and designing the Tokyo subway system \cite{Sun:2016tr,Whiting:2016wq, Adamatzky:2014wg,  Tero:2010va}.  In order to illustrate the novelty of the {\it Physarum Steiner algorithm} as well as the benefits of its use, we will considering the following:
 \begin{itemize} 
 \item \textit{\textbf{Network design}}. We use the algorithm to develop a road network in the United States in Section \ref{roads}. 
  \item \textit{\textbf{Obstacle-avoidance.}} We  use the algorithm to solve the obstacle-avoiding Euclidean Steiner tree problem in Section \ref{obj}. 
 \item \textit{\textbf{Topological surfaces.}} We discuss the algorithm's  adaptability to varying surfaces and boundaries by considering topological surfaces such as the sphere, torus, Klein bottle, and $\mathbb{RP}^2$ in Section \ref{topo}.
 \end{itemize}

\subsection{Road networks}\label{roads}
\begin{figure}[H]
\centering \includegraphics[scale=0.12]{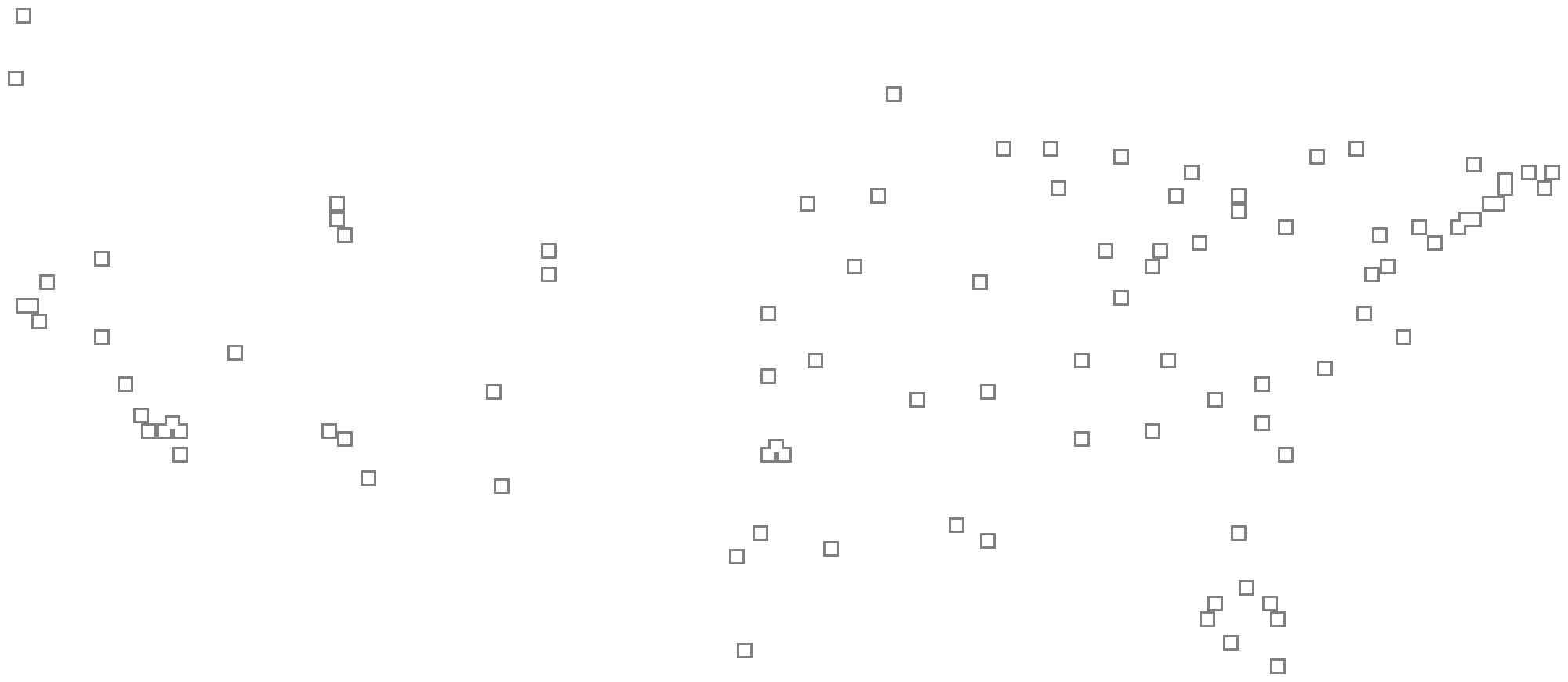}
\caption{The 100 largest cities in the lower 48 states represented as active zones.}
\label{fig:US_points}
\end{figure}

\begin{figure}[H]
\centering \includegraphics[scale=0.14]{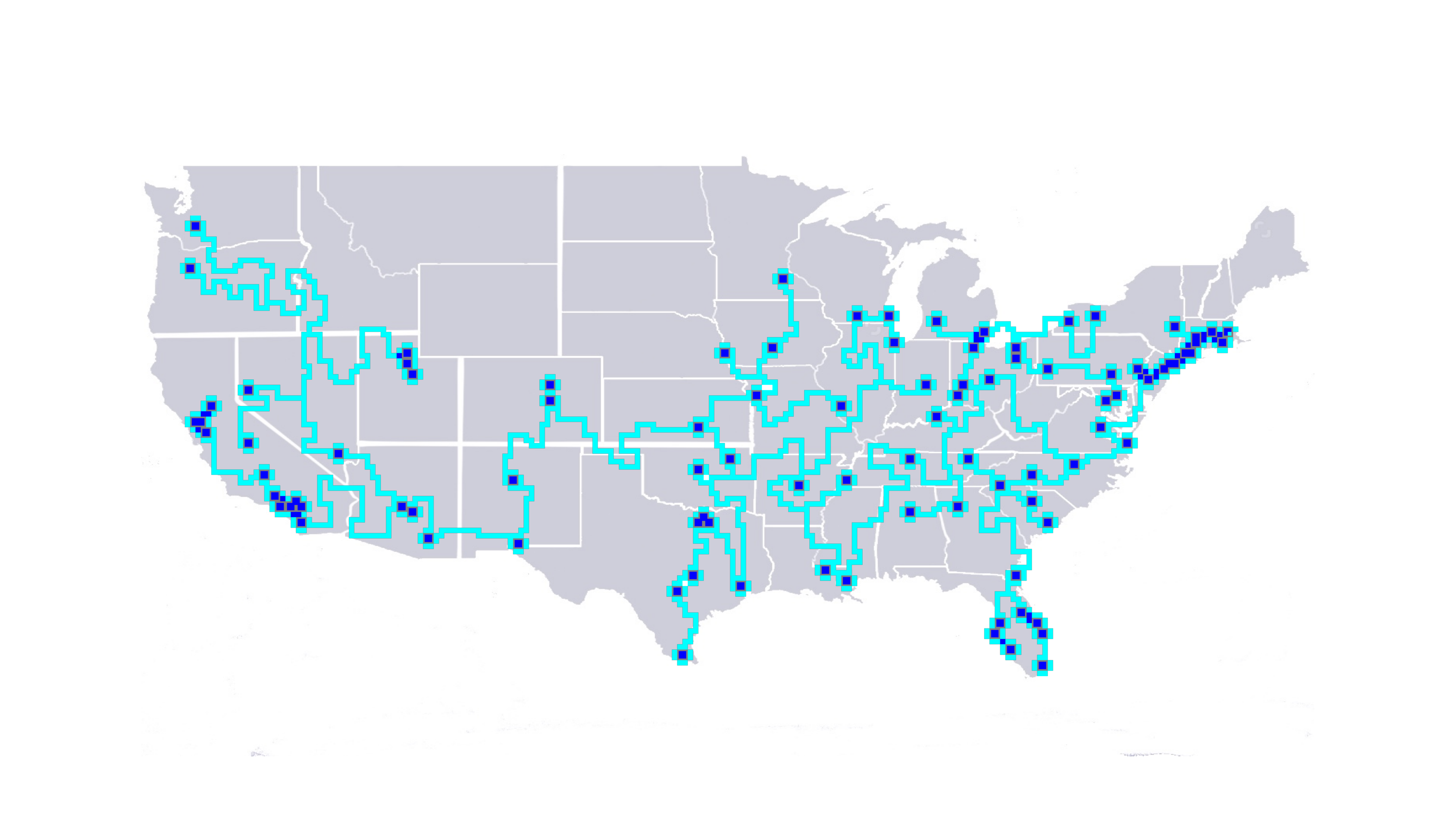}
\caption{The final road network generated by the algorithm. Grey United States map created in Procreate.}
\label{fig:US_MAP}
\end{figure}

The {\it Physarum Steiner algorithm} can be used to build a road network between the largest one hundred cities in the lower 48 United States (excluding Alaska and Hawaii). We use data \cite{simplemaps:ug} of the longitude and latitude of the top 100 cities to generate a rectangular grid of active zones as shown in Figure \ref{fig:US_points}.
We then spawn diamonds of size 7 with a spacing of 1 as shown in Figure \ref{fig:good_diamonds}. After many iterations, the final road network is shown in Figure \ref{fig:US_MAP}.
In addition, the algorithm is particularly suited to the problem of designing transportation systems as before forming a tree, the algorithm's network will have loops that need to slowly be removed. As a result, depending on how much connectivity is wanted, networks the algorithm generated earlier before forming a tree can be used.

\begin{figure}[H]
\centering \includegraphics[scale=0.14]{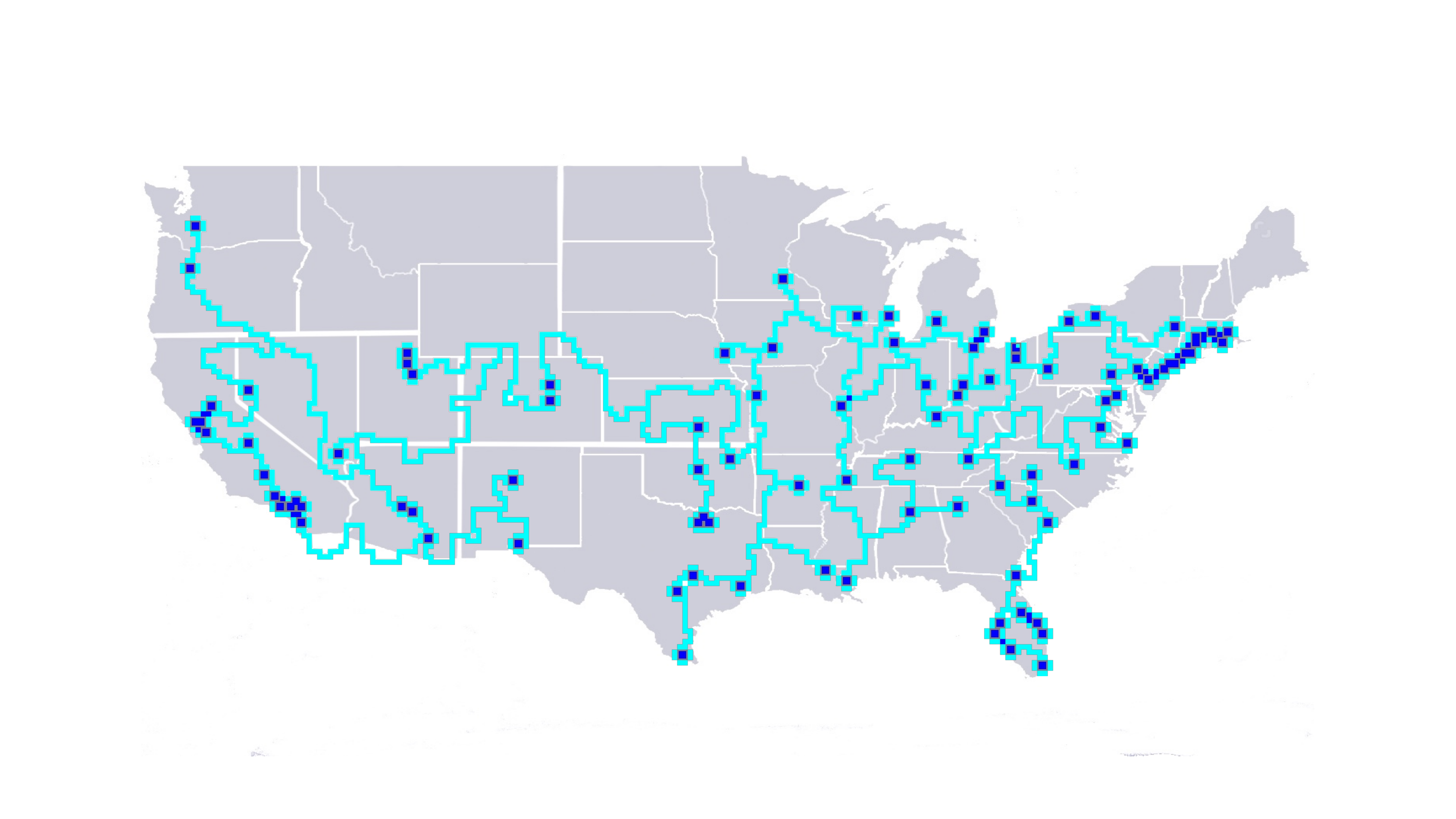}
\caption{Road network generated by the algorithm. Grey United States map created in Procreate.}
\label{fig:US_MAP_LOOPS}
\end{figure}

 For example, in Figure \ref{fig:US_MAP_LOOPS}, we have a network that still contains loops in highly popular areas like the California cities. If we allow the algorithm to continue running, we will get networks with less loops and eventually a tree.

\par We believe that this algorithm can be applied to many similar problems such as designing fiber optic or electric cable networks. We also see VLSI chip design \cite{Cho:2001tz} as an application particularly suited to the algorithm. Due to the algorithm's usage of a square grid, it is easily applied to find rectilinear networks like those required by chips. In addition, due to its time complexity, it should scale well to the large problem of chip design.

 \begin{figure*}[t]
\centerline{
\hspace{-7in} 
\includegraphics[scale=0.105]{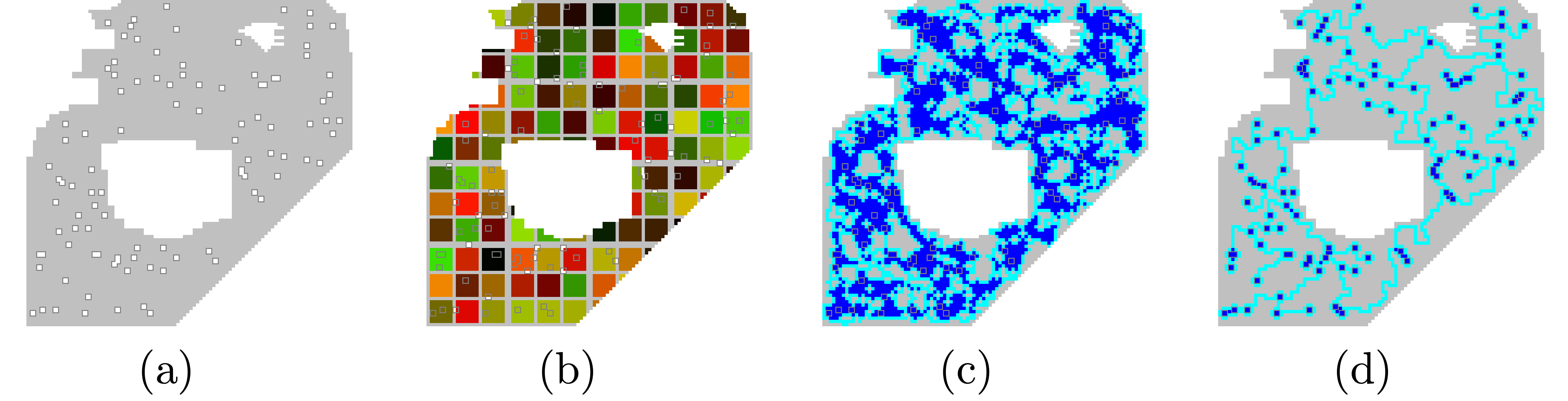} }
\caption{{\bf (a).} Sample boundary map. Grey area is search area and small white squares are points; {\bf (b)}. Initially spawned {\it Physarum} swarm; {\bf (c)}. Algorithm at the end of the foraging stage;  {\bf (d)}. The final network.}   \label{fig:boundary_a}\label{fig:boundary_solution}
\end{figure*}

\subsection{Obstacle avoidance}\label{obj}

\par Due to the cellular automaton nature of this algorithm, it is straightforward to define boundaries or other obstacles that need to be avoided. This is very useful for cases where certain areas need to be avoided such as a lake or the boundary of a country. In addition, we believe that this algorithm is competitive with the current standard obstacle-avoiding Euclidean Steiner algorithm \cite{Zachariasen:1999tv}, which takes multiple hours for graphs with only 150 points. The run time of the {\it Physarum} obstacle avoidance algorithm can be no worse than that discussed in Section \ref{time-complexity}. 

As an example, consider the area in Figure \ref{fig:boundary_a} (a). Here, the grey area represents the search area and the 100 white squares outlined in dark grey are the points. There are many possible real world situations similar to this. For example, the grey area could be a county and all the points represent homes that subscribe to a certain Internet service provider. The big white area in the center could be a lake and the smaller white area could be a dog park. The ISP company could then utilize the {\it Physarum Steiner algorithm} to find networks to lay fiber optic cables as shown in Figure \ref{fig:boundary_solution} (d). This solution was generated in 300,000 iterations and less than 30 seconds.

We begin by spawning a {\it Physarum} swarm of square CELLs of size 7 in Figure \ref{fig:boundary_solution} (b). The CELLs then being to fuse, share intelligence, and find all the points. We choose a solution that still has some loops to increase reliability and ease of future modification to the network.

%

 \subsection{Topological surfaces}\label{topo}
\begin{figure}[H]
\centering \includegraphics[align=c, scale=0.17]{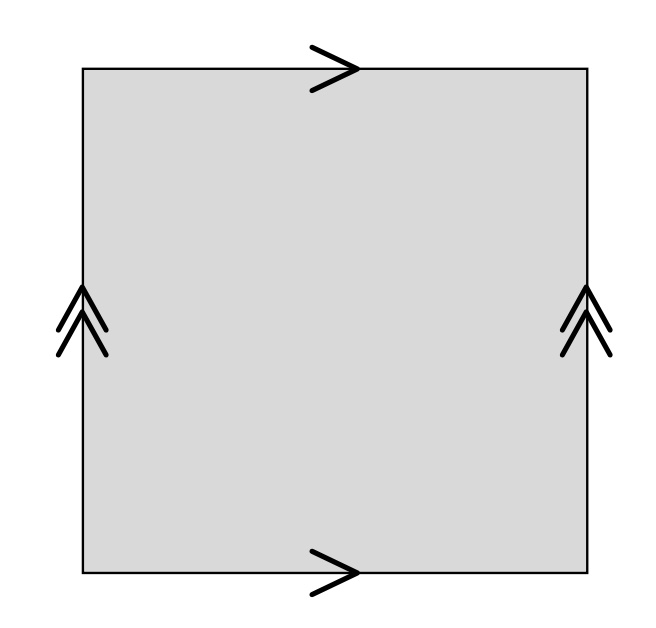} \hspace{3mm} \includegraphics[align=c, scale=0.11]{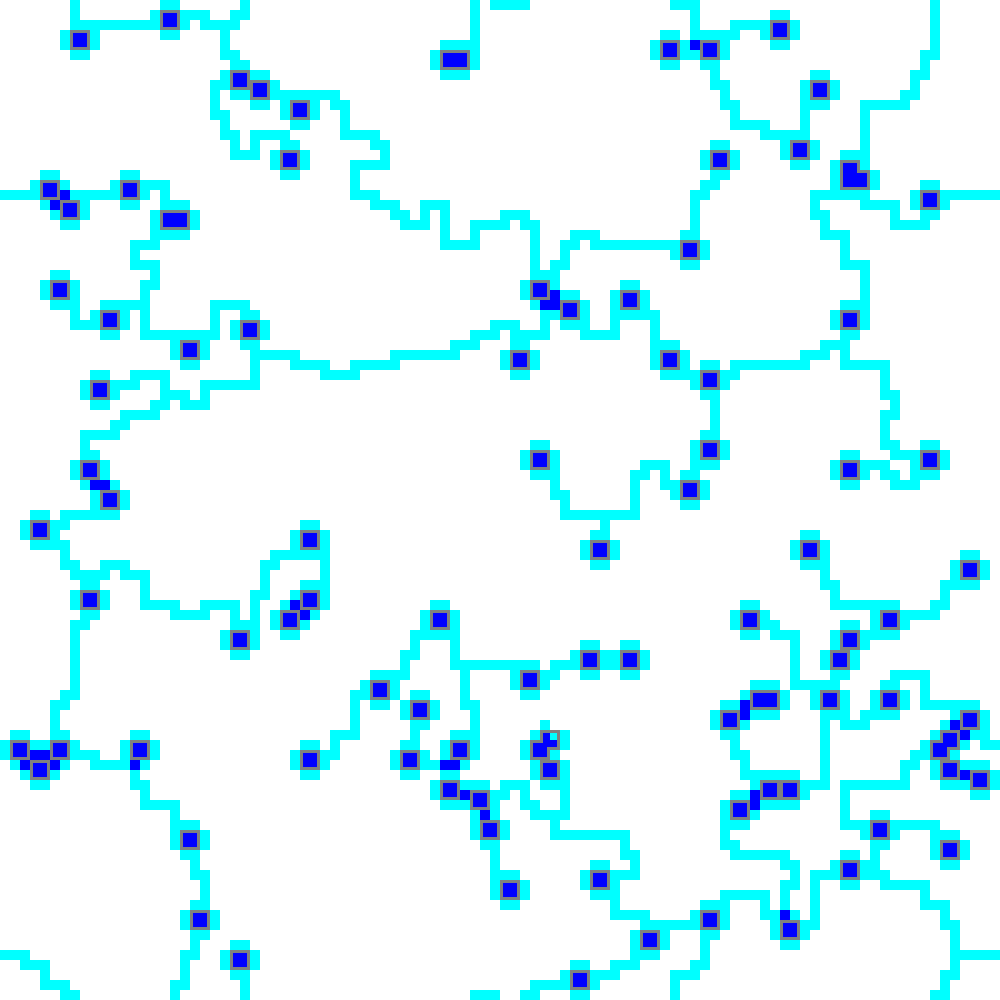}
\caption{Left: identification space of a torus, Right: Steiner tree on identification space of torus}
\label{fig:torus}
\end{figure}

\begin{figure}[H]
\centering \includegraphics[align=c, scale=0.27]{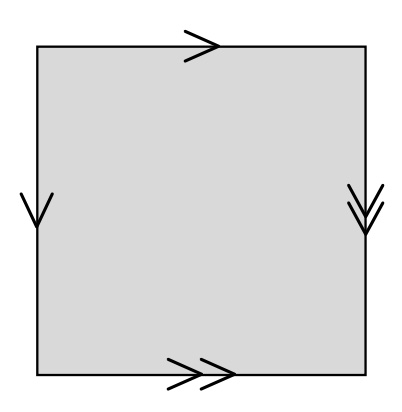} \hspace{3mm} \includegraphics[align=c, scale=0.11]{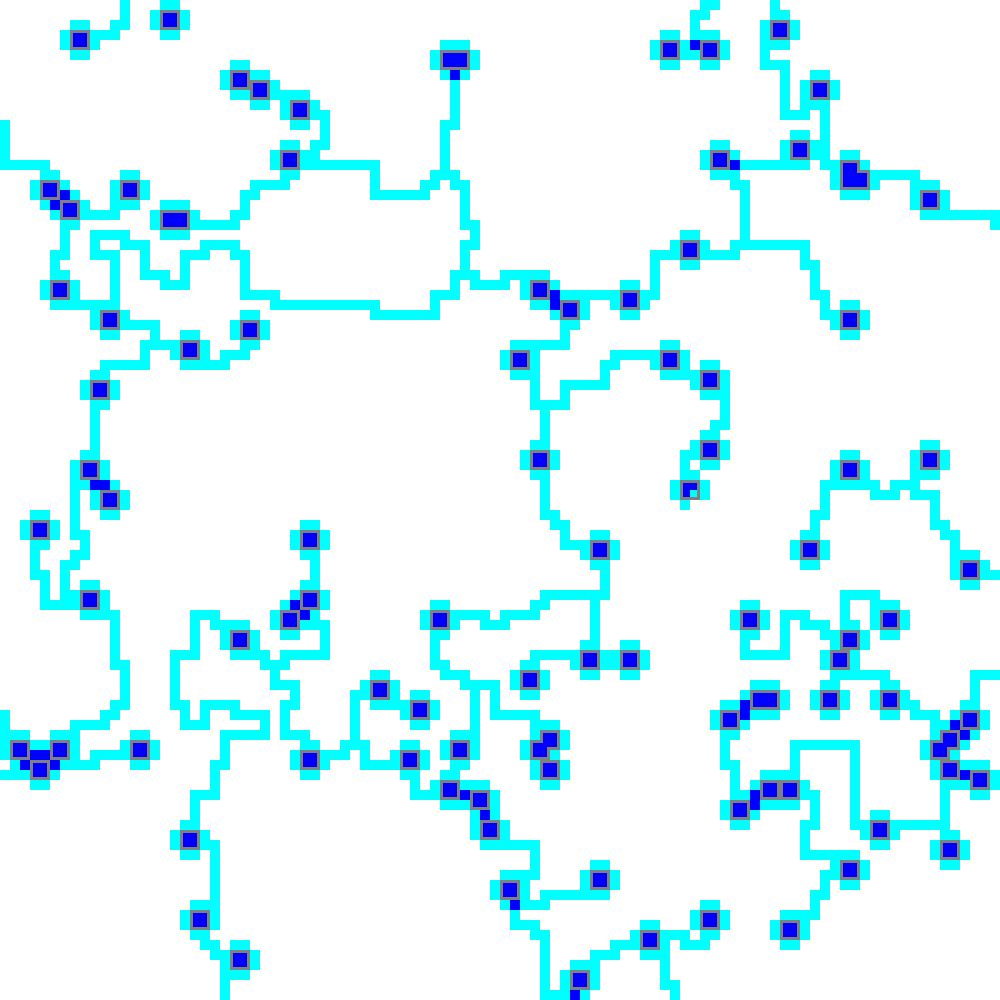}
\caption{Left: identification space of a sphere, Right: Steiner tree on identification space of sphere}
\label{fig:sphere}
\end{figure}

\begin{figure}[H]
\centering \includegraphics[align=c, scale=0.17]{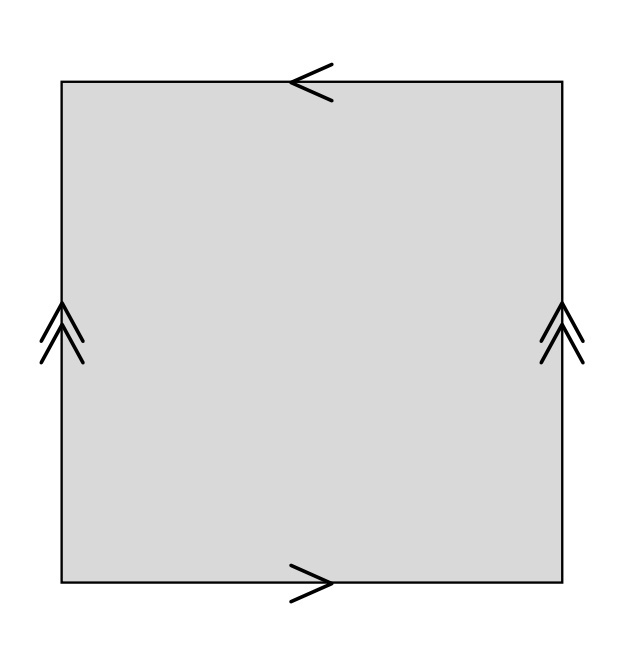} \hspace{3mm} \includegraphics[align=c, scale=0.11]{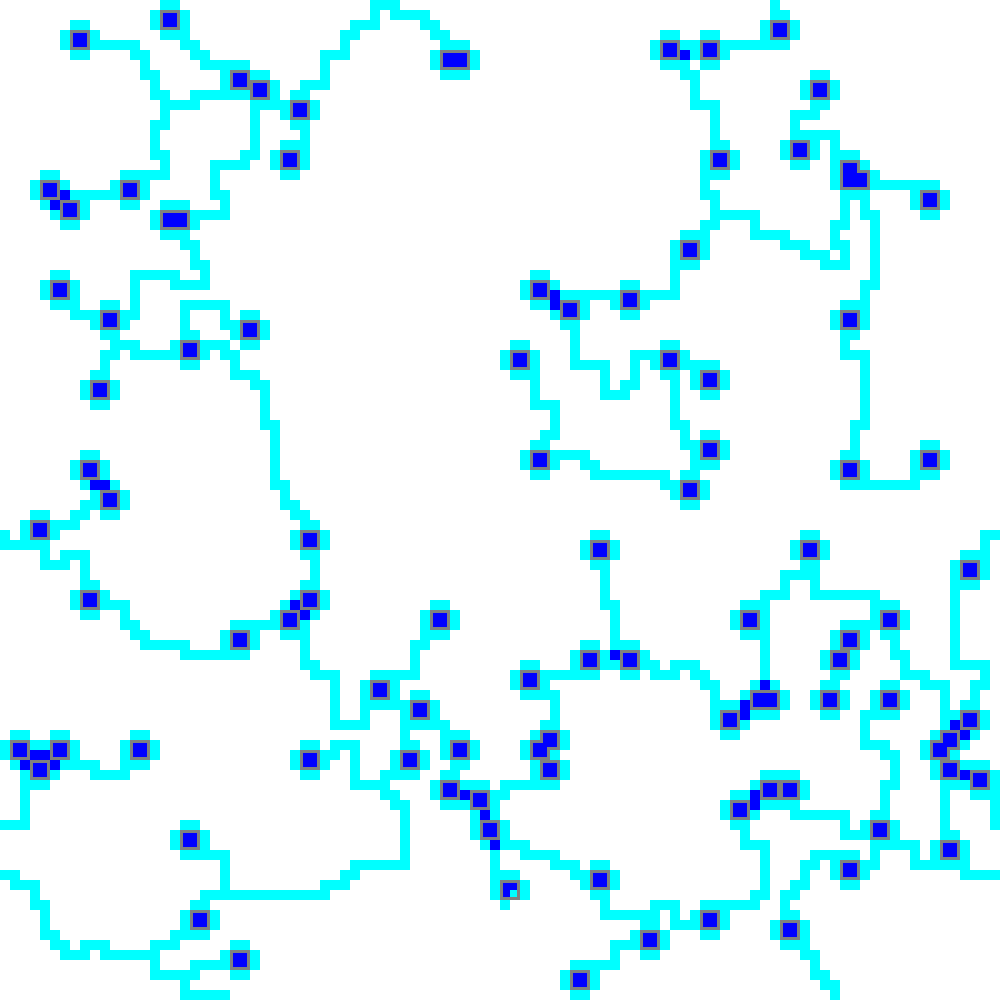}
\caption{Left: identification space of the Klein Bottle, Right: Steiner tree on identification space of a Klein Bottle.}
\label{fig:klein}
\end{figure}

\begin{figure}[H]
\centering \includegraphics[align=c, scale=0.17]{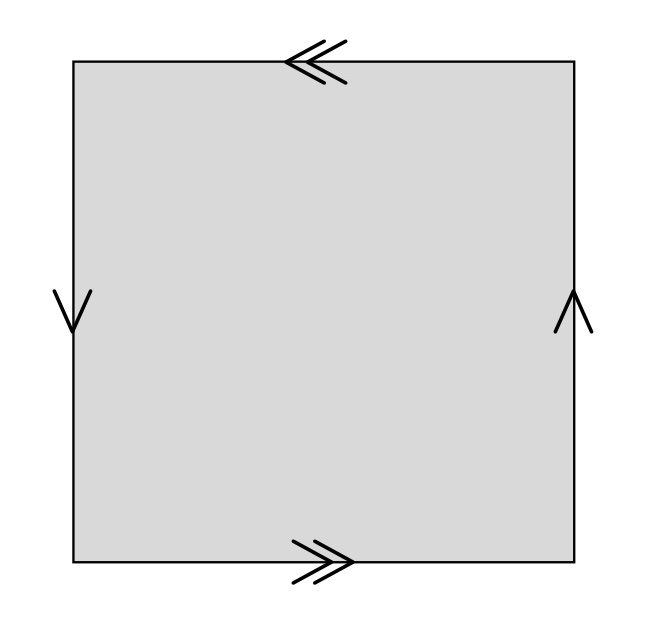} \hspace{3mm} \includegraphics[align=c, scale=0.11]{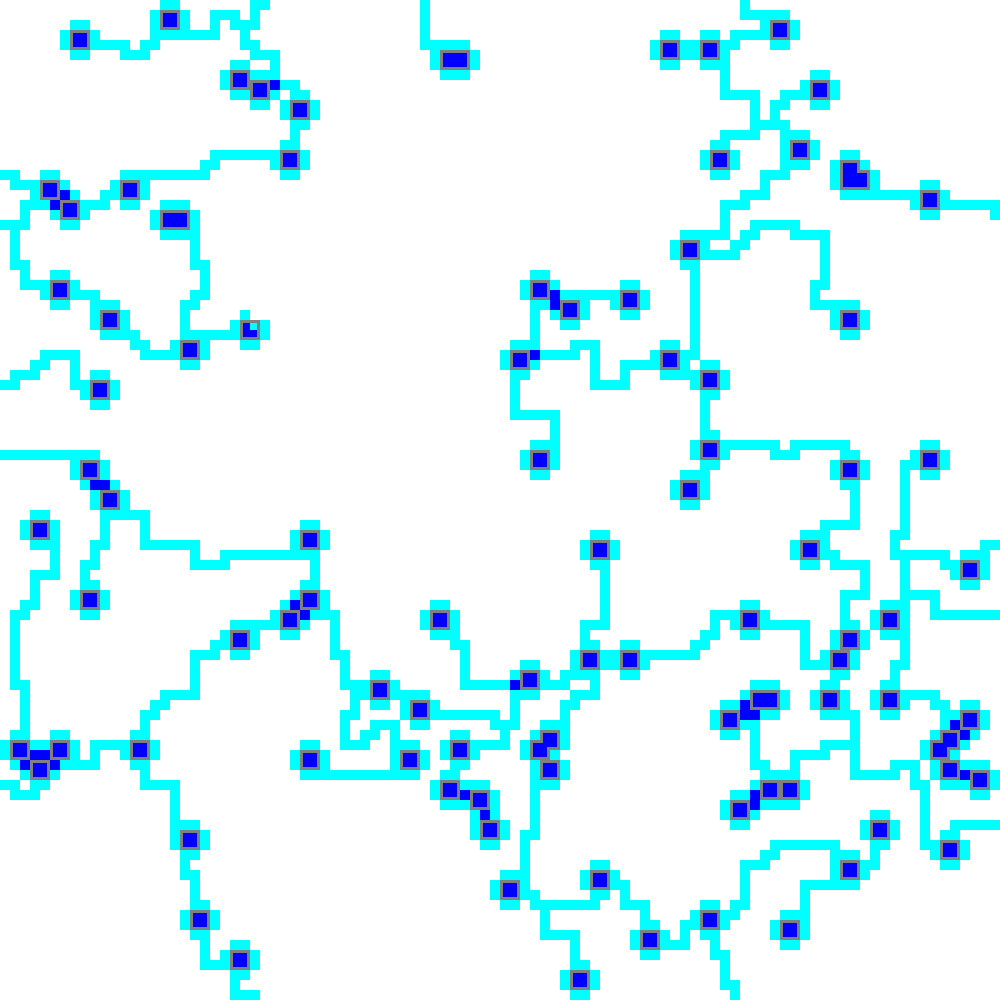}
\caption{Left: identification space of $\mathbb{RP}^2$, Right: Steiner tree on identification space of  $\mathbb{RP}^2$.}
\label{fig:rp2}
\end{figure}

Finally, the {\it Physarum Steiner algorithm} is easily applicable to finding Steiner trees on other topological surfaces. Given the nature of the algorithm, we are able to map coordinates on one edge to the other.  We are able to use square identification spaces to find Steiner trees on the torus in Figure \ref{fig:torus}, the sphere in Figure \ref{fig:sphere}, the Klein bottle in Figure \ref{fig:klein}, and $\mathbb{RP}^2$ in Figure \ref{fig:rp2}. Although we did not do so in this paper, it should be relatively straightforward to modify the algorithm to work on triangular or hexagonal ID spaces, which would enable the modeling of   topological surfaces such as the Dunce cap.



\section{Concluding Remarks} \label{conclusion}
In the present paper we have developed a model of {\it Physarum} swarms and used said swarms to solve the Euclidean Steiner tree problem. We then analyzed the effects of the shape of the cells and the number of cells on the algorithm's performance before discussing the time complexity. Finally, we considered different applications of the   {\it Physarum Steiner algorithm} such as to form road networks, to solve the obstacle-avoiding Steiner problem,  and to understand different topological surfaces.\\
\bigskip

\noindent {\bf Summary of results.}
We present a very unique and different approach to the Steiner problem which we hope will inspire further growth and innovation. 
\bigskip

The current approach to the Euclidean Steiner problem was developed in 1985 so we hope that the new {\it Physarum Steiner algorithm} will spark new ideas in regards to this problem. In addition, the {\it Physarum} swarm is a unique addition to the many swarm algorithms due to its modeling of a single cell rather than a complex organism as well as {\it Physarum}'s ability to share intelligence through fusion. 

\bigskip

There are also some particularly noteworthy advantages of the algorithm.  The algorithm has the ability to incrementally find Steiner trees. The first solution often times contains many loops which are slowly removed with more iterations. This is particularly useful for some applications such as road or cable networks where some degree of redundant connectivity is desired. Due to the algorithm taking place on a grid of squares, the algorithm is particularly applicable to rectilinear Steiner problems such as VLSI design. 
\par In addition, the algorithm performs well on the obstacle-avoidance Euclidean Steiner tree problem. Due to the grid nature of the algorithm, is easily adaptable to different search boundaries. 
\bigskip
\bigskip

\begin{figure}[H]
\centering \includegraphics[scale=0.11]{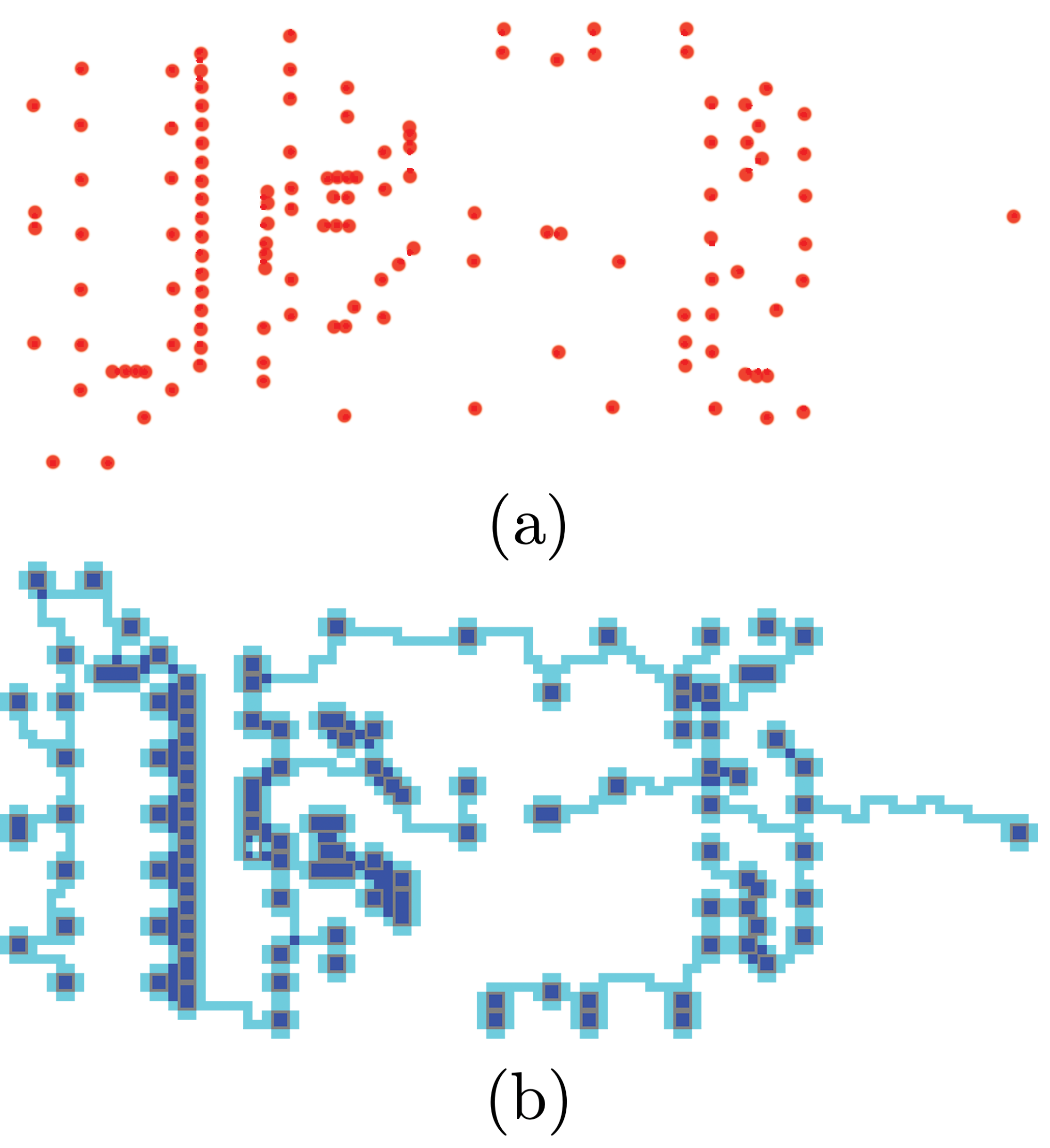}
\caption{On the right is the image provided by \cite{Institute:wn} for their 131-point VLSI dataset. On the right is the tree the {\it Physarum Steiner algorithm} found between the points.}
\label{fig:example}
\end{figure}
  \noindent {\bf Future work.} The new algorithm developed in this paper could be considered to understand the  rectilinear Steiner tree problem, especially considering the importance of the rectilinear Steiner tree to VLSI design. Using data from \cite{Institute:wn}, we lay the groundwork for future VLSI applications.

 In Figure \ref{fig:example}, we consider a set of pads that need to be connected. We represent the pads as active zones and generate a tree between them. In the future, additional improvements would need to be added so that multiple trees can be constructed since many electronic routing problems will contain different signals that need to be routed. In addition, functionality would need to be added to account for multiple layers which is common in pcb design. This could perhaps be done by modeling {\it Physarum} in a 3D perspective.
 
\par On the algorithmic side, future work can be done on trying to optimize the removal of loops. Currently, the algorithm takes a very long time to shrink and remove loops since bubbles tend to get stuck in active zones instead of stopping in the middle of a loop. In addition, more work can be done on decreasing the time complexity of each iteration. There has already been some work done on this such as using a Disjoint Set Union to track CELLs, but there is still the potential for further improvement. In addition, future work can be done in finding the actual runtime instead of true iterations, which would provide better comparison against other Steiner tree algorithm. Multiple implementations of the {\it Physarum Steiner algorithm} have been made publicly available \cite{Hsu:wo}.
\par From the biological perspective, future work includes physically growing multiple {\it Physarum} organisms and seeing if they fuse and form Steiner trees in a method similar to the {\it Physarum Steiner algorithm}.


\bigskip
 \noindent {\bf Acknowledgments.} The  authors   are thankful to MIT
PRIMES-USA for the opportunity to conduct this research together. The authors would also like to thank Fidel A. Schaposnik for bringing   \textit{Physarum Polycephalum} to our attention after it started growing in his boiler. The authors would also like to thank Rafe Mazzeo for his advice on the topological surface application of this algorithm.
  The work of Laura Schaposnik is partially supported through the NSF grant  CAREER DMS 1749013.   \\
 
 \noindent {\bf Affiliations.}\\
  (a)  Valley Christian High School, San Jose,   USA. \\
  (b) Institut des Hautes Etudes Scientifiques, France. \\
  (c)  University of Illinois, Chicago,  USA. \\
  
  \pagebreak
  
\addcontentsline{toc}{section}{%
\protect\numberline{9}%
Bibliography}%
%
\bibliographystyle{IEEEtran}

\bibliography{IEEEabrv,YauBib}
\end{document}